\documentclass[11pt,document,nofootinbib,superscriptaddress,onecolumn,preprintnumbers,balancelastpage]{article}

\pdfoutput=1
\usepackage{jheppub}

\usepackage[section]{placeins}
\usepackage{graphicx,color}
\usepackage{amsmath,amssymb,amsfonts}
\usepackage{stackrel}
\usepackage{enumitem}
\usepackage[english]{babel}
\usepackage{verbatim}
\usepackage{hyperref}
\usepackage{comment}
\usepackage[normalem]{ulem}
\usepackage{xcolor}
\usepackage{subcaption}
\usepackage{multirow, float}
\usepackage{dcolumn}
\usepackage{physics}
\usepackage[bottom]{footmisc}
\usepackage[toc,page]{appendix}
\usepackage{bm}
 
\graphicspath{
  {./}
  {figures/}
}

\newcommand{\tf}{\texorpdfstring}
\newcommand{\gev}{~\text{GeV}}
\newcommand{\tev}{~\text{TeV}}
\newcommand{\mev}{~\text{MeV}}

\newcommand{\onbb}{0\nu\beta\beta}

\newcommand{\gsim}{\buildrel > \over {_\sim}}
\def\nn{\nonumber}

\newcommand{\fb}{~\text{fb}}

\newcommand{\abi}{~\text{ab}^{-1}}
\newcommand{\fbi}{~\text{fb}^{-1}}
\newcommand{\yri}{~\text{year}^{-1}}

\definecolor{byzantine}{rgb}{0.74, 0.2, 0.64}

\newcount\Comments  
\newcommand{\kibitz}[2]{\ifnum\Comments=1\textcolor{#1}{#2}\fi}

\preprint{}

\makeatletter
\gdef\@fpheader{}
\makeatother

\title{Uncovering a chirally suppressed mechanism of $0\nu\beta\beta$ decay with LHC searches}

\author[a]{Michael~L.~Graesser,}
\author[b,c]{Gang~Li,}
\author[d,c,e]{Michael~J.~Ramsey-Musolf,}
\author[c]{Tianyang~Shen,}
\author[c]{Sebasti\'an Urrutia-Quiroga}

\affiliation[a]{\footnotesize Theoretical Division, Los Alamos National Laboratory, Los Alamos, NM 87545, USA}
\affiliation[b]{\footnotesize School of Physics and Astronomy, Sun Yat-sen University, Zhuhai 519082, P.R. China}
\affiliation[c]{\footnotesize Amherst Center for Fundamental
    Interactions, Department of Physics, University of Massachusetts,
    Amherst, MA 01003, USA}
\affiliation[d]{\footnotesize Tsung-Dao Lee Institute and School of
    Physics and Astronomy, Shanghai Jiao Tong University, 800
    Dongchuan Road, Shanghai, 200240 China }
\affiliation[e]{\footnotesize Kellogg Radiation Laboratory,
    California Institute of Technology, Pasadena, CA 91125, USA}

\emailAdd{michaelgraesser@gmail.com}
\emailAdd{ligang65@mail.sysu.edu.cn}
\emailAdd{mjrm@sjtu.edu.cn, mjrm@physics.umass.edu}
\emailAdd{tysimonshen@gmail.com}
\emailAdd{surrutiaquir@umass.edu}

\preprint{ACFI-T22-02, LA-UR-21-32454}

\abstract{
$\Delta L =2$ lepton number violation (LNV) at the TeV scale could provide an alternative interpretation of positive signal(s) in future neutrinoless double beta $(0\nu\beta\beta)$ decay experiments. An interesting class of models from this point of 
view are those that at low energies give rise to dimension-9 vector operators and a 
dimension-7 operator, both of whose $0\nu\beta\beta$-decay rates are ``chirally suppressed".
We study and compare
the sensitivities of $0\nu\beta\beta$-decay experiments and LHC searches to a simplified model in this class  of TeV-scale LNV that is also $SU(2)_L \times U(1)_Y$ gauge invariant. 
The searches for $0\nu\beta\beta$ decay, which are here diluted by a chiral suppression of the vector operators, 
are found to be less constraining than LHC searches whose reach is  increased by the assumed kinematic accessibility of the mediator particles.
For the chirally suppressed dimension-7 operator generated by TeV-scale mediators, in contrast, $0\nu\beta\beta$-decay searches place strong constraints on the size of the new Yukawa coupling. Signals of this model at the LHC and $0\nu\beta\beta$-decay experiments are entirely uncorrelated with the observed neutrinos masses, as these new sources of LNV give negligible contributions to the latter.
We find the prospects for the high-luminosity LHC and ton-scale $0\nu\beta\beta$-decay experiments to uncover the chirally-suppressed mechanism with TeV-scale LNV to be promising. 
We also comment on the sensitivity of the $\onbb$-decay lifetime to certain unknown low-energy constants that in the case of dimension-9 {\it scalar} operators are expected to be large due to non-perturbative renormalization. 
}

\begin{document}
\emergencystretch 3em 
\maketitle
\newpage

\section{Introduction}

In the Standard Model (SM) of particle physics, lepton number is conserved in all perturbative interactions. The well-known seesaw mechanism~\cite{Minkowski:1977sc,Gell-Mann:1979vob,Glashow:1979nm,Yanagida:1980xy,Mohapatra:1979ia,Schechter:1980gr} that explains the origin of tiny neutrino masses strongly motivates the existence of overall lepton number violating interactions in extensions to the SM, which could appear at the TeV scale, in for example the type-II seesaw mechanism~\cite{Schechter:1980gr,Abada:2007ux}. There have been extensive studies of TeV-scale LNV at intensity and energy frontiers, inspired by the non-zero neutrino masses~\cite{Angel:2012ug}, baryon asymmetry of the Universe~\cite{Pilaftsis:2003gt,Deppisch:2017ecm,Harz:2021psp}, and experimental accessibility~\cite{Helo:2013dla,Peng:2015haa,DeGouvea:2019wnq}.

We particularly concentrate on the violation of lepton number by two units, $\Delta L =2$, which is related to the possible Majorana nature of neutrinos~\cite{Schechter:1980gr} and can be tested directly in the process of neutrinoless double beta $(\onbb)$ decay.
In this context, it is significant that current cosmological observations 
constrain the sum of the light neutrinos masses to $\lesssim 0.12$ eV \cite{Planck:2018vyg}, while future cosmological 
surveys may completely exclude the so-called inverted hierarchy~\cite{Hazumi:2012gjy,Euclid:2014mgp, Brinckmann:2018owf,Abazajian:2019eic, DESI:2019jxc}. 
Despite the impressive limits on $\onbb$-decay lifetimes, it may prove challenging for 
the next-generation $\onbb$-decay experiments, which will probe the full range of the inverted hierarchy, 
to detect the conventional Majorana neutrino mass mechanism~\cite{Li:2020flq}. That potentially gloomy 
outcome may actually be invigorating: a negative outcome from cosmological observations excluding the inverted hierarchy
coupled with a future positive 
$\onbb$-decay detection necessarily prompts us to investigate alternatives to Majorana neutrino masses that are responsible for $\onbb$ decay.

The systematic application of chiral perturbation theory $(\chi$PT$)$ and chiral effective field theory $(\chi$EFT$)$ to $0 \nu \beta \beta$ decay was first pioneered by Ref.~\cite{Prezeau:2003xn}, which also pointed out general the dominance of pionic contributions
for several dimension-9 operators.
\footnote{In the case of specific models (R-parity-violating supersymmetry) the importance of pion interactions was first noticed in Ref.~\cite{Faessler:1996ph}.} 
In that work, the classification of possible $\Delta L =2$ LNV operators at the dimension-9 level, including only quark and lepton fields, that contribute to $\onbb$ decay was performed using Weinberg power counting \cite{Weinberg:1990rz,Weinberg:1991um}.
\footnote{This terminology refers to Weinberg's prescription 
for organizing the momentum expansion in multi-nucleon scattering processes. A cogent summary can be found in Ref.~\cite{Kaplan:1996xu}.}
In particular, the quark-lepton operators are organized according to the power counting of the mapped hadron-lepton operators in chiral effective field theory.
These $\onbb$-decay operators do not necessarily generate the observed neutrino masses, and could give sizable contributions to $\onbb$ decay~\cite{Rodejohann:2011mu}. In the framework of simplified models outlined in~\cite{Bonnet:2012kh}, the potential for the Large Hadron Collider (LHC) to probe TeV-scale LNV is promising~\cite{Helo:2013ika,Helo:2013dla,Peng:2015haa}, as LHC searches can be complementary to $\onbb$-decay experiments. 

In this work, we will  study the phenomenology of TeV-scale LNV arising from certain dimension-9 vector operators discussed below, 
using a simplified model that provides an ultra-violet (UV) completion for these operators and that is $SU(2)_L \times U(1)_Y$ gauge invariant. 
The use of a simplified model allows us to assess the relative sensitivities of $\onbb$ decay and searches for LNV in high energy $pp$ collisions, as the relevant partonic center of mass energies in the latter may not justify integrating out the new particles responsible for the low-energy effective operators.  Importantly, at low-energies, the leading contribution of the  dimension-9 operators to $\onbb$ decay occurs at next-to-next-to-leading order (N$^2$LO) in Weinberg's power counting, 
in striking contrast to most other dimension-9 LNV  operators (so-called ``scalar'' operators) that can occur at leading order (LO). Moreover, a renormalization group analysis of the vector operators implies that their treatment according to Weinberg power counting is robust \cite{Cirigliano:2018yza, Cirigliano:2019vdj}; the promotion of higher order contact operators to LO that enters the scalar channel (as well as the conventional light Majorana neutrino exchange mechanism)
 \cite{Cirigliano:2018hja,Cirigliano:2019vdj,Cirigliano:2020dmx,Cirigliano:2021qko} does not occur in this case. 
 
 It is worth noting that the interplay between $\onbb$-decay experiments and LHC searches in a similar simplified model has been studied previously in Ref.~\cite{Helo:2013ika}. \footnote{Specifically, 
 `topology I' with the `decomposition 2-ii-b' in the terminology of the diagrams and Table I of that reference.}
There are notable differences however between that work and the present analysis. The simplified model studied here is $SU(2)_L \times U(1)_Y$ gauge invariant, 
 whereas the models and topologies studied in Ref.~\cite{Helo:2013ika} are only $SU(3)_c \times U(1)_{em}$ invariant. 
Electroweak invariance has consequential difference, as we show it:
 relates the dimension-9 vector operators that contribute to $0 \nu \beta \beta$ decay to other $\Delta L=$2 operators, that generate neutrino masses at three-loop level; and makes for a more complex and richer LHC phenomenology, since for instance, different particles in the same 
 electroweak doublet produce different final states upon decay. 
 In our validation and recasting of existing LHC analyses, we perform parton showering, jet reconstruction, and use a fast detector simulation (\texttt{Delphes3}~\cite{deFavereau:2013fsa}).  A subsequent analysis comparing the LHC reach of a model 
 that generates a {\it scalar} LNV operator to the analysis of Ref.~\cite{Helo:2013ika}, 
 found that important Standard Model and detector backgrounds were overlooked \cite{Peng:2015haa}. In performing projections for the high-luminosity LHC, here we instead re-scale existing LHC analyses assuming the signal-to-background ratio remains the same.
 In our $0 \nu \beta \beta$ analysis 
 we include here the perturbative QCD one-loop renormalization group evolution of the LNV operators from the TeV scale down to the hadron scale.
 Moreover, the analysis of the $0 \nu \beta \beta$-decay process itself 
 performed here uses a complete basis for dimension-9 vector operators \cite{Graesser:2016bpz}, 
 and makes use of more recent results using chiral effective field theory \cite{Cirigliano:2017djv, Cirigliano:2018yza}. Ref.~\cite{Helo:2013ika} appears to not include so-called 4-quark `color-octet' operators, and only includes the contribution of the 4-nucleon $(np) (np)$ interaction, but ignores the pion contribution \cite{Peng:2015haa} even though it occurs at the same order in the chiral power counting. Although these contributions appear with unknown low-energy constants, we consider their impact on the $0 \nu \beta \beta$-decay rate by varying the low-energy constants over an $\mathcal{O}(1)$ range.

This model also generates a dimension-7 LNV operator, that makes a long-distance 
contribution to $\onbb$ decay. Here however, its LO contribution to 
$\onbb$ decay vanishes, and the next-leading contribution is proportional to the electron mass or outgoing electron 
energies. While the $\onbb$-decay amplitude is suppressed, the corresponding $\onbb$-decay bound on the Yukawa coupling is still quite strong, suggesting this interaction -- which does not contribute to the dimension-9 vector 
operators -- is not relevant at LHC energies.

While current and future ton-scale $\onbb$ decay experiments are insensitive to the details of the underlying LNV mechanism
because they are only sensitive to the decay rate, and not for example, angular or energy correlations between the two 
outgoing electrons, LHC searches have a greater potential to uncover it, if the associated mass scale is at the TeV. 
The reason is that if the particles that generate these operators have masses close to the TeV mass scale, they 
could be produced on-shell at the LHC, greatly increasing their production cross section. 
In the case of dimension-9 LNV {\em scalar} operators this potential has already been demonstrated \cite{Helo:2013ika} \cite{Peng:2015haa} \cite{Harz:2021psp}. 

Turning to $\onbb$ decay, because these vector operators contribute to the amplitude at N$^2$LO, their contributions to the $\onbb$-decay rate are na\"ively suppressed by a factor of $(\Lambda_\chi/m_\pi)^{4}\sim 2\times 10^3$, with $\Lambda_\chi \simeq 1$ GeV a typical hadronic scale and $m_\pi$ the pion mass. 
As a result of this suppression -- which is a purely a consequence of low-energy hadronic physics -- and the increase of production cross section with on-shell particles,  LHC 
searches are relatively stronger at constraining these operators than direct $\onbb$-decay limits.

It is noted that Refs.~\cite{Helo:2013ika} \cite{Harz:2021psp} have studied the sensitivities of
$0 \nu \beta \beta$-decay and LHC
searches, and also found that LHC searches can have greater reach. 
As aforementioned, this is caused by the increase in production cross section arising from the mediators being kinematically accessible at the LHC.  Here, we emphasize that in case of dimension-9 LNV vector operators, the chiral suppression of the $\onbb$-decay rate will further promote the sensitivity of LHC searches in comparison with $\onbb$-decay searches. In contrast, 
the perturbative QCD renormalization group evolution of the vector operators is found to 
cause their Wilson coefficients at the hadronic scale to be slightly enhanced compared to the TeV scale, 
partly 
offsetting the previous effect. These points are discussed in more quantitative detail in Section \ref{sec:model_DBD-dim9}.

We derive the sensitivities of $\onbb$ decay for the KamLAND-Zen~\cite{KamLAND-Zen:2016pfg} and future ton-scale experiments, and investigate the current and projected sensitivities to TeV-scale LNV at the LHC in our simplified model. Our results clearly illustrate the complementary role of these experiments in probing the TeV-scale physics described here.

One might be concerned that LNV at the TeV scale necessarily generates too large of a neutrino mass, 
and that that constraint renders such new phenomena unobservable at current and near-future LHC and 
$\onbb$-decay experiments. This concern is model-dependent, and here 
we find that while these new sources of LNV do indeed generate neutrino masses, they 
 are induced at three-loop level or higher level in the case of dimension-9 vector 
 operators, and for the dimension-7 operator at two-loop level \footnote{Ref.~\cite{Helo:2013ika} notes 
 that because of the Schechter-Valle ``black box" theorem~\cite{Schechter:1981bd}, 
 a neutrino mass is generated at four-loop level. However, because of the $SU(2)_L$ invariance of this model, 
 neutrino masses are generated at three-loop level, and at a value larger -- significantly larger than a loop factor -- than given 
 by the theorem. This result is not in contradiction with the Schecter-Valle ``black box'' theorem, for recall it only provides a {\it minimum} estimated value to 
 the neutrino mass given a positive detection of a $0 \nu \beta \beta$-decay lifetime. These and other 
 aspects are discussed further in 
 Sections \ref{sec:neutrino-masses-dim9} and \ref{sec:neutrino-masses-dim7}.}. In both 
 circumstances the induced neutrino masses are further suppressed by powers of the light quark 
 and electron masses and are entirely negligible.
This does mean that the model discussed here cannot explain the origin of 
neutrino masses.

A potential shortcoming of the simplified model described here is -- in the absence of imposing any additional symmetries --  the presence of large flavor-changing neutral currents and charged lepton flavor violating processes. As the interactions most relevant for our analysis here involve only first generation leptons and quarks, we consider only the flavor-diagonal components of the model and defer a study of the flavor- non-diagonal interactions to future work.
Indeed,
our aim is to 
demonstrate the important input
the LHC can provide towards solving what will may hopefully be a future $\onbb$-decay \lq\lq inverse problem".

The rest of this paper is organized as follows. In Sec.~\ref{sec:model_DBD}, we introduce a simplified model containing leptoquarks that, at low energies, generates dimension-9 LNV vector operators
and a dimension-7 LNV operator. In Sec.~\ref{sec:model_DBD-dim9} and 
Sec.~\ref{sec:model_DBD-dim7} we then calculate the half-lives of $0\nu\beta\beta$ decay arising 
from the dimension-9 and dimension-7 operators, and estimate their contributions to 
the neutrino mass in Sec.~\ref{sec:neutrino-masses-dim9} and Sec.~\ref{sec:neutrino-masses-dim7}, respectively. Then, in Sec.~\ref{sec:LHC}, we recast the existing LHC searches  and make projections for future prospects. In Sec.~\ref{sec:results}, we discuss the sensitivities of the LHC and $\onbb$-decay experiments to TeV-scale LNV in our  model. 
In Appendices~\ref{app:scalar-vs-vector}, \ref{app:neutrino_mass}, and \ref{app:prod_decay}, we extend the discussion about hadron-level $0\nu\beta\beta$ amplitudes, show how the neutrino mass arises at three-loop level in the model, 
and provide all the relevant decay widths of the new particles, respectively.
We conclude in Sec.~\ref{sec:conclusion}.

\section{Model and \tf{$\onbb$}{0vbb} decay }
\label{sec:model_DBD}

To see how the $\onbb$ decay and collider probes interplay with each other in the tests of TeV-scale LNV, we will work in the context of simplified models. 
The systematic decomposition of $\onbb$-decay operators with the possible realizations are sketched in Refs.~\cite{Bonnet:2012kh,Chen:2021rcv}. We will particularly pay attention to the UV realization of a class of $\onbb$-decay operators which are suppressed at hadron-level in chiral power counting, but are potentially accessible at colliders and next-generation $\onbb$-decay experiments and give small contributions to the neutrino masses. 

Consider the following model with a scalar field $S\in (1,2)_{1/2}$, a leptoquark field $R\in (3,2)_{1/6}$ and a Dirac fermion field $\Psi\in (1,2)_{-1/2}$,
where $(X,Y)_{Z}$ corresponds to the representations under the $SU(3)_C$, $SU(2)_L$ and $U(1)_Y$ gauge groups, respectively. 
They interact with the SM fields via
\begin{align}
\label{eq:interactions}
\mathcal{L}_{\text{int}} &=y_{qd} \bar{Q} S d_R + y_{qu} \bar{u}_R S^T  \epsilon Q + y_{e\Psi} \bar{e}_R S^\dagger \Psi_L \nn \\
& + \lambda_{ed} \bar{L}\epsilon R^*d_R + \lambda_{u\Psi} \bar{\Psi}_R R u_R^c  +\lambda_{d\Psi} \epsilon \bar{\Psi}_L R^* d_R \nn \\
& + y_{e \Psi}^\prime \bar{\Psi}_L H e_R +\text{h.c.}\;, 
\end{align}
where $Q=(u,d)_L^T$, $L=(\nu_{e}, e)_L^T$, $\epsilon=i\sigma^2$,  $\psi_{L/R}\equiv P_{L/R} \psi$, $u_R^c = P_L C \bar{u}^T$ with $P_{L/R} \equiv (1\mp \gamma_5)/2$, $\psi$ denoting the fermion fields and $C$ being the usual charge conjugation matrix, and ``$\text{h.c.}$'' represents the Hermitian conjugate terms. 
The fields $S$ and $\Psi_L$ have the same quantum numbers as the SM Higgs and lepton doublet fields, respectively.
A mass term $\overline{\Psi}_R L$ can be removed by a field redefinition of $\Psi_L$ and $L$.
Following Ref.~\cite{Wise:2014oea}, we assume for simplicity 
that at tree-level $S$ has no vacuum expectation value and so at this order all of its components are physical.

With three generations of SM fermions there is no reason {\it a priori} 
for the Yukawa couplings of $S$ to quarks and charged leptons to be aligned with the SM Yukawa 
couplings. 
Tree-level exchange of $S$ generically leads to dangerous flavor-changing 
neutral current processes and charged lepton number violating processes. We will not address 
this problem here, since our focus is on illustrating the interplay between the LHC and $\onbb$-decay
experiments.

This model violates overall lepton number 
whenever $\lambda_{ed} \lambda_{u\Psi}\neq 0$ or $\lambda_{d\Psi} \lambda_{u\Psi}\neq 0$. To see that, consider a fictitious lepton number $U(1)_L$, 
under which the fields are charged, and the Yukawa couplings are treated as spurions 
and also charged to make the Lagrangian invariant. Thus the lepton numbers $q(L)=q(e)=q(\Psi)=1$, 
$q(S)=0$, and $q(R)=r$ can be arbitrary. Then the 
Yukawa couplings necessarily have a charge such that $q(\lambda_{ed})=q(\lambda_{d\Psi})=1+r$ and $q(\lambda_{u\Psi})=1-r$,
implying
$q(\lambda_{ed} \lambda_{u\Psi})=q(\lambda_{d\Psi} \lambda_{u\Psi})=2$.
The term in Eq.~\eqref{eq:interactions}, $\lambda_{d\Psi} \epsilon \bar{\Psi}_L R^* d_R$,
does not contribute to $\onbb$ decay at tree-level and 
is not considered further.

\begin{figure}[!htb] 
\centering
\includegraphics[width=0.5\textwidth]{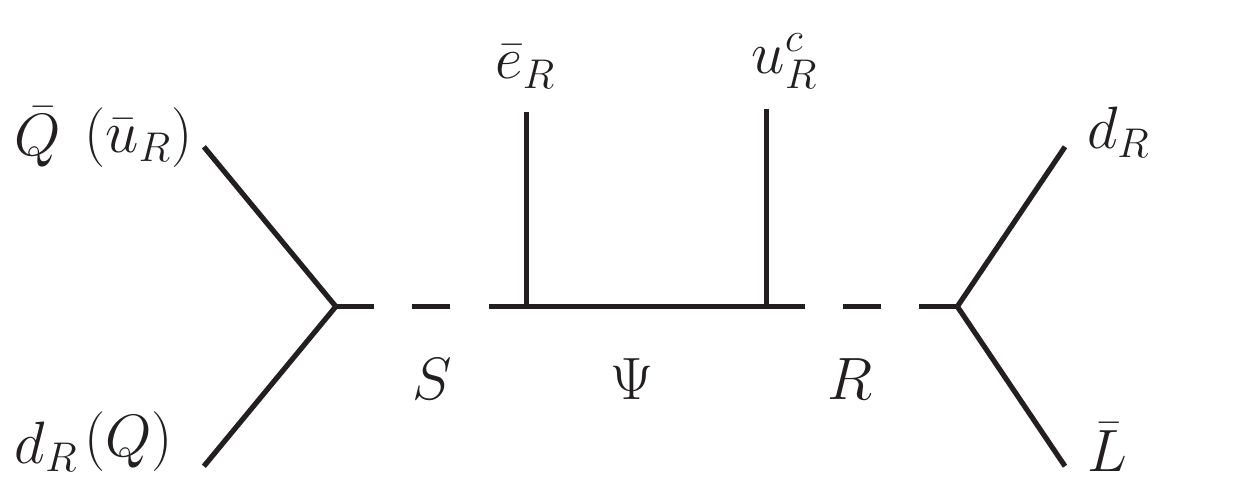}	
\caption{Quark-level Feynman diagrams that induce $\Delta L=2$ vector $\onbb$-decay operators at low energies. 
}
\label{fig:feyn_DBD_quark}
\end{figure}

At low energies this model generates 
dimension-9 and dimension-7 operators that contribute to $\onbb$ decay. In each of 
these cases the amplitude for $\onbb$ decay is suppressed, either by $m^2_\pi$ in the case of the 
dimension-9 vector operators, or in the case of the dimension-7 operator, 
by the electron mass or energies 
of the outgoing electrons.
We next discuss these 
in turn.

\subsection{Dimension-9 operators}
\label{sec:model_DBD-dim9}

At the quark level, the interactions in this model lead to two $\onbb$-decay contributions shown in Fig.~\ref{fig:feyn_DBD_quark}, arising from the interactions in Eq.~\eqref{eq:interactions} proportional to 
$y_{qd}$ and $y_{qu}$, respectively. After integrating out the $S$, ${R}$, and $\Psi$ fields, the effective interactions take the forms of 
\begin{align}
(\bar{u}_L d_R) (\bar{e}_R u_R^c) (\bar{e}_L d_R ) \;,\quad (\bar{u}_R d_L) (\bar{e}_R u_R^c) (\bar{e}_L d_R )\;
\end{align}
with $u_R^c\equiv (u_R)^c$.
Using the Fierz identity,
we obtain
\begin{align}
(\bar{e}_R u_R^c) (\bar{e}_L d_R ) = \dfrac{1}{2} (\bar{u}_R \gamma^\mu d_R) (\bar{e}_R \gamma_\mu e_L^c)\;.
\end{align}

In the approach of the SM effective field theory (SMEFT), the effective interactions after integrating out the heavy fields are~\cite{Prezeau:2003xn,Graesser:2016bpz,Cirigliano:2017djv,Cirigliano:2018yza,Liao:2020jmn,Li:2020xlh} 
\begin{align}
\label{eq:SMEFT}
\mathcal{L}_{{\rm SMEFT}}^{(9)} = \dfrac{1}{\Lambda^5}&(C_{Qu1} O_{Qu1} + C_{Qd1} O_{Qd1}  ) + \text{h.c.}\;,
\end{align}
where $\Lambda$ is the  LNV scale, and the dimension-9 quark-lepton vector operators are expressed as
\begin{align}
\label{eq:dimension-9_SMEFT}
\begin{split}
O_{Qu1} &= (\bar{ u}_R Q) (\bar{u}_R\gamma_\mu d_R)(\bar{e}_R \gamma^\mu L^{c})\;,\\
O_{Qd1} &=\epsilon_{ij} (\bar{Q}^i d_R) (\bar{u}_R\gamma_\mu d_R)(\bar{e}_R \gamma^\mu L^{jc})\;.
\end{split}
\end{align}
The Wilson coefficients are given by 
\begin{align}
\begin{split}
\label{eq:UV-Wilson-coefficients}
\dfrac{C_{Qu1}}{\Lambda^5} &=\dfrac{y_{e\Psi}\lambda_{ed}\lambda_{u\Psi}}{2m_S^2 m_R^2 m_\Psi}y_{qu}\;,\\
\dfrac{C_{Qd1}}{\Lambda^5} &=  \dfrac{y_{e\Psi}\lambda_{ed}\lambda_{u\Psi}}{2m_S^2 m_R^2 m_\Psi}y_{qd}\;,
\end{split}
\end{align}
where $m_R$, $m_S$, and $m_\Psi$ are physical masses of $R$, $S$ and $\Psi$, respectively.

Note that under the fictitious $U(1)_L$ introduced earlier, the charge of these Wilson coefficients is 
exactly cancelled by the charge of the effective operators $O_{Qu1}$ or $O_{Qd1}$, so 
that the above $\mathcal{L}_{{\rm SMEFT}}^{(9)}$ is invariant. 
Hereafter we assume without loss of generality that all of the couplings are real.

We evolve the operators $O_{Qu1}$ and $O_{Qd1}$ down to the electroweak scale, which mix with the color octet vector operators $O_{Qu2}$ and $O_{Qd2}$, respectively, written as
\begin{align}
\begin{split}
O_{Qu2} &= (\bar{ u}_R t^a Q) (\bar{u}_R t^a \gamma_\mu d_R)(\bar{e}_R \gamma^\mu L^{c})\;,\\
O_{Qd2} &=\epsilon_{ij} (\bar{Q}^i t^a d_R) (\bar{u}_R t^a \gamma_\mu d_R)(\bar{e}_R \gamma^\mu L^{jc})\;.
\end{split}
\end{align}
The corresponding Wilson coefficients $C_{Qu2}$ and $C_{Qd2}$ are normalized in a similar way as $C_{Qu1}$ and $C_{Qd1}$ in the Lagrangian. The one-loop QCD RG equation for the Wilson coefficients $C_{Qu1}$ and $C_{Qu2}$ is~\cite{Cirigliano:2018yza},
\begin{align}
\label{eq:RGE}
\dfrac{d}{d\ln \mu}
\begin{pmatrix}
C_{Qu1} \\
C_{Qu2}
\end{pmatrix}
=
\dfrac{\alpha_s}{4\pi} 
\begin{pmatrix}
-\dfrac{40}{9} & \dfrac{80}{27} \\[6pt]
\dfrac{4}{3} & \dfrac{46}{9}
\end{pmatrix}
\begin{pmatrix}
C_{Qu1} \\
C_{Qu2}
\end{pmatrix}\;.
\end{align}
where $\alpha_s$ is the strong coupling. The evolution of the Wilson coefficients $C_{Qd1}$ and $C_{Qd2}$ is govern by the RG equation of the same form with the substitution: $C_{Qu1}\to C_{Qd1}$ and $C_{Qu2}\to C_{Qd2}$.

Below the electroweak scale, the effective interactions are written in the low-energy effective field theory (LEFT)~\cite{Cirigliano:2018yza,Prezeau:2003xn,Graesser:2016bpz,Li:2020tsi} 
\begin{align}
\label{eq:dimension-9_lag}
\mathcal{L}_{\text{LEFT}}^{(9)} = \dfrac{1}{v^5} \sum_{i=6}^9 C_i^\prime O_i^{\mu\prime} \bar{e}\gamma_\mu \gamma_5 e^c + \text{h.c.}\;,
\end{align}
where the vacuum expectation value $v=246\gev$, and the  vector operators are given by
\begin{align}
\label{eq:dimension-9_operators}
\begin{split}
O_6^{\mu \prime}&= (\bar{q}_R \tau^+ \gamma^\mu q_R ) (\bar{q}_R \tau^+  q_L)\;,\\
O_7^{\mu \prime}&= (\bar{q}_R t^a \tau^+ \gamma^\mu q_R ) (\bar{q}_R t^a \tau^+  q_L)\;,\\
O_8^{\mu \prime}&= (\bar{q}_R \tau^+ \gamma^\mu q_R ) (\bar{q}_L \tau^+  q_R)\;,\\
O_9^{\mu \prime}&= (\bar{q}_R t^a \tau^+ \gamma^\mu q_R ) (\bar{q}_L t^a \tau^+  q_R)\;,
\end{split}
\end{align}
where $\tau^+=(\tau^1+i \tau^2)/2$ with $\tau^1$ and $\tau^2$ being the Pauli matrices, and $q_{L,R}=(u,d)^T_{L,R}$ denote the left-handed and right-handed quark isospin doublets. For the color octect operators $O_7^{\mu \prime}$ and $O_9^{\mu \prime}$, $t^a=\lambda^a/2$, $a=1,\ldots, 8$, $\hbox{Tr} [t^a t^b]= \delta^{ab}/2$, $\lambda^a $ denote the $SU(3)$ 
Gell-Mann matrices in the fundamental representation, and the summation over the index $a$ is assumed. 

The matching conditions at the electroweak scale $m_W=80.4\gev$ are
\begin{align}
\begin{split}
C_{6(7)}^{ \prime}(m_W) &=  \dfrac{v^5}{2\Lambda^5} C_{Qu1(2)}(m_W) \;,\\
C_{8(9)}^{ \prime}(m_W) &= \dfrac{v^5}{2\Lambda^5} C_{Qd1(2)}(m_W)\;.
\end{split}
\end{align}
The RGEs for the Wilson coefficients in the LEFT take the same form in Eq.~\eqref{eq:RGE} with the substitution: $C_{Qu1}\to C_6^\prime (C_7^\prime)$, $C_{Qu2}\to C_8^\prime (C_9^\prime)$; and with the change in 
the $\beta$-function of $\alpha_s$. 
Choosing a typical high scale~\footnote{We neglect the difference if $m_{S,R,\Psi}$ deviate from $2\tev$, which is a minor effect. } $\Lambda=2\tev$, the  Wilson coefficients $C_{6,7,8,9}^\prime$ at the scale $m_0=2\gev$,  are given by
\begin{align}
\label{eqn:RGeqn1}
\begin{pmatrix}
C_{6(8)}^\prime (m_0) \\
C_{7(9)}^\prime (m_0)
\end{pmatrix} =
\dfrac{\bm{Z}}{2} \left[ \frac{v^5}{\Lambda^5} \right]
\begin{pmatrix}
C_{Qu1(Qd1)} (\Lambda) \\
C_{Qu2(Qd2)} (\Lambda)
\end{pmatrix}\;
\end{align}
with 
\begin{align}
\label{eqn:RGeqn2}
\bm{Z} = 
\begin{pmatrix}
1.43  & -0.23 \\
-0.10 & 0.68
\end{pmatrix}\;.
\end{align}
The matrix ${\bm Z}$ gives the one-loop running from the scale $\mu=\Lambda$, where  $C_{Qu1}(\Lambda)$ and $C_{Qd1}(\Lambda)$ are given in Eq.~\eqref{eq:UV-Wilson-coefficients} and $C_{Qu2}(\Lambda)=C_{Qd2}(\Lambda)=0$, to the scale $\mu=m_0$.

\begin{figure}[!htb] 
\centering
\includegraphics[width=0.35\textwidth]{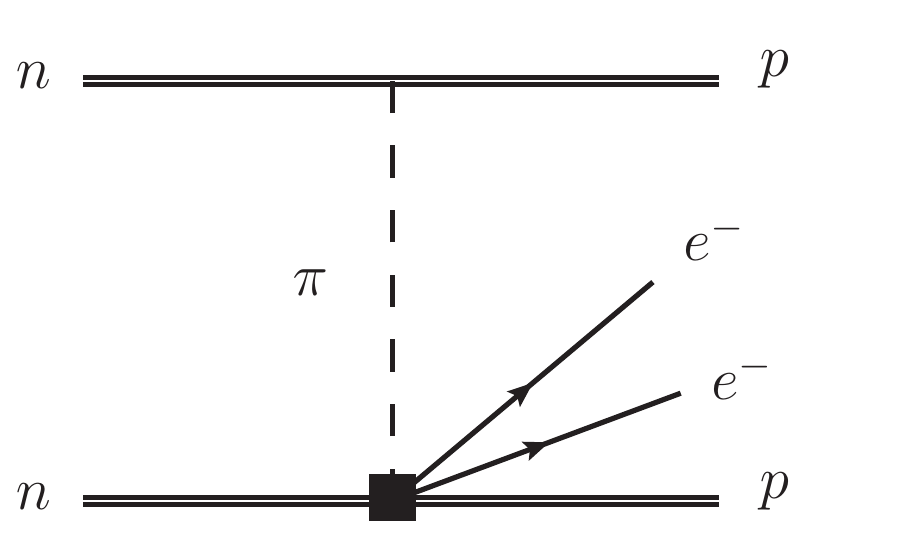}	
\includegraphics[width=0.35\textwidth]{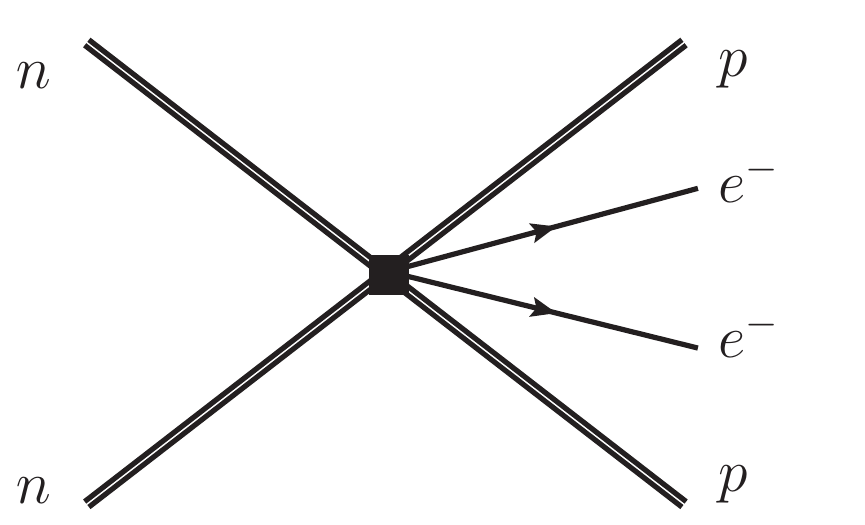}	
\caption{
Hadron-level Feynman diagrams 
for $\onbb$ decay
induced by the LNV operators $\pi NN ee$ (left) and $NNNNee$ (right), which are denoted by the black squares. 
}
\label{fig:feyn_DBD_hadron}
\end{figure}

In chiral perturbation theory the quark-lepton operators are mapped onto hadron-lepton operators defined 
below the scale $\Lambda_\chi\sim 1\gev$, with the non-perturbative QCD dynamics encoded by low-energy constants (LECs)~\cite{Bernard:1995dp,Gasser:1983yg}. 

\begin{figure}[!htb] 
\centering
\includegraphics[width=0.6\linewidth]{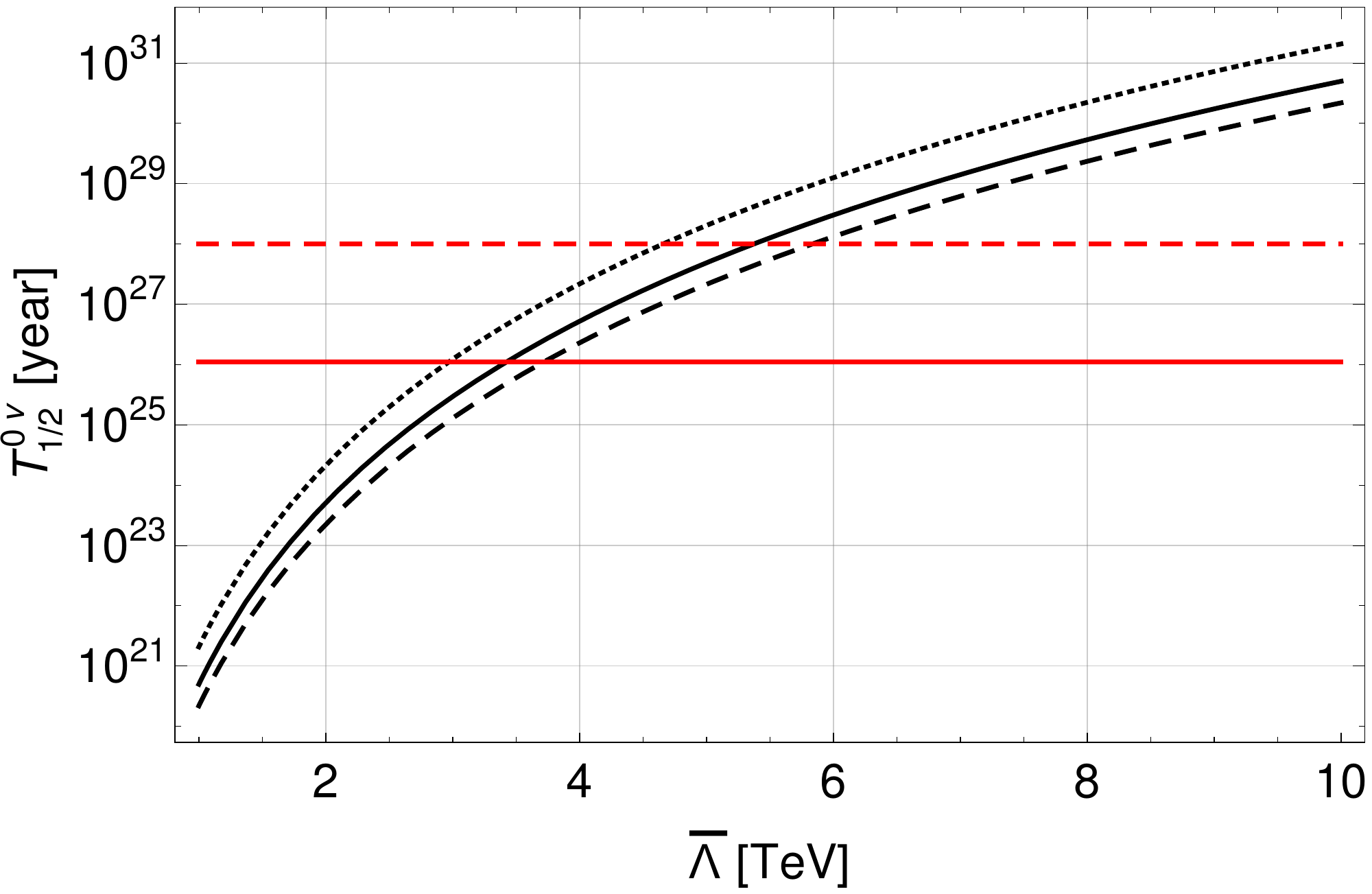}	
\caption{Half-lives of $\onbb$ decay as a function of the effective LNV scale $\overline\Lambda\equiv \Lambda/\left|C_{Qu1}+C_{Qd1}\right|^{1/5}$. 
The solid black curve is obtained with $g_V^{\pi N}=\tilde g_V^{\pi N} =1$ and $g^{NN}_6=g^{NN}_7=0$. The dotted (dashed) black curve is obtained with the same settings of $g_V^{\pi N}$, $\tilde g_V^{\pi N}$ but with $g^{NN}_6=g^{NN}_7=1/2$ $(-1/2)$. The solid red and dashed red lines correspond to the constraints on the half-life in KamLAND-Zen and future ton-scale experiments for $^{136}$Xe. 
} 
\label{fig:DBD-sensitivity}
\end{figure}

In several ways the behavior of the vector operators in chiral perturbation theory 
is in striking contrast to that of scalar operators. For vector operators, their 
lowest order local interaction with pions is of the form
$(\pi \partial_\mu \pi) \overline{e} \gamma^\mu \gamma_5 e^c$ which, after using the pion equations of 
motion, is $\pi \pi ee$, but importantly suppressed
by a factor of $m_e$ and negligible \cite{Prezeau:2003xn,Graesser:2016bpz} - for further details see Appendix \ref{app:scalar-vs-vector}. Unlike scalar operators, 
vector operators do not induce any operators at LO in the chiral power counting.
The leading contribution of the 
vector operators $O_{6,7,8,9}^{\mu\prime}\bar{e} \gamma_\mu \gamma_5 e^c$ to the $\onbb$-decay 
amplitude instead arises from their mapping onto the hadronic operators 
$\pi NN ee$ and $NNNNee$ ~\cite{Prezeau:2003xn,Graesser:2016bpz,Cirigliano:2018yza}
which are at NLO and N$^2$LO in Weinberg's power counting~\cite{Weinberg:1990rz,Weinberg:1991um}, respectively. 
While requiring that the amplitude is 
regulator independent implies that the LECs for these two operators are related, in the case of vector operators, no ``promotion'' 
of the $NNNNee$ operators to lower chiral order is needed, and at least for these vector operators, 
Weinberg's power counting appears correct \cite{Cirigliano:2018yza,Cirigliano:2019vdj}. In the case of 
vector operators then, the $\pi NN ee$ and $NNNNee$ 
operators make comparable 
contributions to the $\onbb$-decay amplitude. The $\onbb$-decay rate however has a sizable  uncertainty 
due to the unknown values of these LECs, as is illustrated in Fig. \ref{fig:DBD-sensitivity}.

In heavy-baryon chiral perturbation theory~\cite{Jenkins:1990jv} the effective 
Lagrangian for the leading hadron-level LNV interactions
$\pi NN ee$ and $NNNNee$
is given by~\cite{Cirigliano:2018yza}
\begin{eqnarray}
\label{eq:eff-Lag-had}
\mathcal{L}^{\rm vector}_{\text{eff}} &=& \mathcal{L}^{\rm vector}_{\pi N}  + \mathcal{L}^{\rm vector}_{N N} ~, \\
\mathcal{L}^{\rm vector}_{\pi N} &=& \frac{1}{v^5}\sqrt{2} g_A F_\pi \bar{p} S^\alpha \cdot \partial_\alpha \pi^- n \Big[ g_V^{\pi N} C_V^{(9)} +\tilde{g}_V^{\pi N} \tilde C_V^{(9)} \Big] \times v^{\mu} \bar{e}\gamma_\mu \gamma_5 e^c ~, \\
\mathcal{L}^{\rm vector}_{NN} &=& \frac{1}{v^5} (\bar{p} n)(\bar{p}n) 
\Big[g_6^{NN} C_V^{(9)}+g_7^{NN} \tilde{C}_V^{(9)} \Big] \times v^\mu \bar{e} \gamma^\mu \gamma_5 e^c  ~,
\end{eqnarray} 
where $F_\pi=91.2\mev$, $g_A=1.27$, and $S^\alpha$ and $v^\mu$ are the nucleon spin and velocity. The coefficients
\begin{align}
C_V^{(9)} &=  C_6^\prime(m_0) + C_8^\prime(m_0) \;,\nn\\
\tilde C_V^{(9)} &= C_7^\prime (m_0) + C_9^\prime (m_0) \;~.
\end{align}
The LECs $g_V^{\pi N},\tilde g_V^{\pi N}, g_6^{NN}, g^{NN}_7= \mathcal{O}(1)$ 
in the na\"ive dimensional analysis (NDA)~\cite{Manohar:1983md,Weinberg:1989dx}, and little is known about them.
Inspecting the numerical solution to the RG equations (\ref{eqn:RGeqn1}) and (\ref{eqn:RGeqn2}), one finds 
$C_V^{(9)}=1.43v^5/(2\Lambda^5) [C_{Qu1} (\Lambda)+C_{Qd1} (\Lambda)]$ -- so that this Wilson coefficient has an $\mathcal O(1)$ enhancement at low scales \footnote{Note that the factor of $v^5/(2\Lambda^5)$ comes from the definitions of Wilson coefficients.} --
and a small non-vanishing $\tilde{C}_V^{(9)}$ has been generated.

In Fig.~\ref{fig:feyn_DBD_hadron} we show the Feynman diagrams for the $\onbb$ decay at hadron-level, that is the  transition $nn\to pp ee$, which are induced by the NLO operators $\pi NN ee$ and N$^2$LO operators $NNNNee$. In chiral power counting, the transition amplitudes of these two diagrams are at the same order.
For the vector operators consider here, the amplitude for nuclear $\onbb$ decay $0^+\to 0^+$ is \cite{Cirigliano:2018yza}
\begin{align}
\mathcal{A}_{\text{vector}} = \dfrac{g_A^2 G_F^2 m_e }{\pi R_A} \mathcal{A}_M \bar{u}(k_1) \gamma_0 \gamma_5 C \bar{u}^T (k_2)\;,
\label{eq:amplitude-vector-operators}
\end{align}
where $k_1$ and $k_2$ are the momenta of the emitted electrons, $G_F$ is Fermi coupling constant, the radius $R_A= 1.2 A^{1/3}$~fm with $A$ the number of nucleons of  nucleus,
and the reduced amplitude is given by 
\begin{eqnarray}
\label{eq:reduced_amp}
\mathcal{A}_M &=& \dfrac{m_\pi^2}{m_e v}\left[ \dfrac{1}{2} (g_V^{\pi N} C_V^{(9)}+\tilde g_V^{\pi N} \tilde C_V^{(9)}) M_{P,sd} 
- \frac{2}{g^2_A} \left(g^{NN}_6 C^{(9)}_V + g^{NN}_7 \tilde{C}^{(9)}_V \right) M_{F,sd} \right]
\end{eqnarray}
with $M_{P,sd}\equiv (M_{GT,sd}^{AP} + M_{T,sd}^{AP})$, where the basic NMEs $M_{GT,sd}^{AP}=-2.80$, $ M_{T,sd}^{AP}=-0.92$ 
and $M_{F,sd}=-1.53$ for $^{136}$Xe calculated using the quasi-particle random phase approximation (QRPA)~\cite{Hyvarinen:2015bda}. The reader is referred to Ref.~\cite{Cirigliano:2017djv,Cirigliano:2018yza} for 
definitions of these basic NMEs.
The inverse half-life of the $\onbb$ decay is expressed as
\begin{align}
\label{eq:half-life_1}
\left(T_{1/2}^{0\nu}\right)^{-1}= g_A^4 G_{09} |\mathcal{A}_M|^2\;,
\end{align}
where the phase space factor $G_{09}=2.8\times 10^{-14}\yri$ for $^{136}$Xe~\cite{Horoi:2017gmj}, following a trivial rescaling as discussed in \cite{Cirigliano:2018yza,Cirigliano:2017djv}; see also Ref.~\cite{Stefanik:2015twa,Doi:1985dx}. Three-body nucleon 
interactions such as $NNNNNNee$ contributing to $0^+ \rightarrow 0^+$ are expected to contribute at higher-order in the Weinberg power counting 
and are not considered here.

As discussed above, the vector operators $O_{6,7,8,9}^{\mu\prime}\bar{e} \gamma_\mu \gamma_5 e^c$ are mapped onto the operators $ \pi NN ee$ and $NNNNee$, whose contributions are suppressed in chiral power counting. This can be seen in  Eq.~\eqref{eq:reduced_amp} with the amplitude $\mathcal{A}_M$ being proportional to $m_\pi^2$, while amplitudes for the LO hadron-lepton operators are proportional to $\Lambda_\chi^2$~\cite{Prezeau:2003xn,Graesser:2016bpz,Cirigliano:2018yza}. An explicit UV completion of these scalar quark-lepton operators can be found in Ref.~\cite{Li:2021fvw}.

As an aside, one may na\"ively expect that, in the absence of any underlying 
model and consequently no linear ordering of the magnitudes of the Wilson coefficients of 
these operators, the vector operators 
are quantitatively less important than {\it scalar} operators in $\onbb$ decay due to the suppression of the $\onbb$-decay amplitude by a factor of $(\Lambda_\chi/m_\pi)^2\sim 60$. However, this level of suppression may not be realized in practice. To illustrate, consider a comparison with 
the $\onbb$-decay amplitude $(\mathcal{A}_{\text{scalar}})$ induced 
by  scalar operators. It turns out that this suppression is sensitive to the size of certain unknown LECs as well as NMEs. 
To see why, first consider the limit where the unknown LECs for operators $NNNNee$  contributing to $\mathcal{A}_{\text{scalar}}$ follow the Weinberg power counting expectation of $\mathcal{O}(1)$, 
then 
\begin{align}
\frac{\mathcal{A}_{\text{vector}}}{\mathcal{A}_{\text{scalar}}} \simeq \frac{m^2_\pi}{m^2_N}\frac{M_{P,sd}}{M_{PS,sd}}~,
\label{eq:amplitude-ratio-1}
\end{align}
where $m_N \sim \Lambda_\chi$ is the nucleon mass and $\mathcal{A}_{\text{scalar}}$ arises from the LO pion-exchange interaction. Here $M_{PS,sd}$ is a NME that appears in the expression for $\mathcal{A}_{\text{scalar}}$; more details on each can be found in Appendix \ref{app:scalar-vs-vector}.
As shown in Appendix \ref{app:scalar-vs-vector}, the ratio of NMEs is about $6-8$,
across a variety of isotopes and methods for estimating their values. Thus, the na\"ively expected suppression may too severe by nearly an order of magnitude. In addition, recent work suggests that the the LECs for the $NNNNee$ 
arising from the scalar operators
may be considerably larger than $\mathcal{O}(1)$~\cite{Cirigliano:2018hja,Cirigliano:2018yza,Cirigliano:2019vdj, Cirigliano:2020dmx, Cirigliano:2021qko} -- for more 
details the reader is referred to Appendix \ref{app:scalar-vs-vector}. 
The $NNNNee$  contribution to $\mathcal{A}_{\text{scalar}}$ involves yet another NME, further clouding the estimated magnitude of the full scalar amplitude.

Substituting the values of the phase space factor, NMEs and the other constants into Eq.~\eqref{eq:half-life_1}, we obtain 
\begin{align}
\label{eq:half-life_2}
(T_{1/2}^{0\nu})^{-1} &\simeq 6.0 
\times 10^{-22}\big[(0.98 g_V^{\pi N} - g_6^{NN})- (0.069 \tilde{g}_V^{\pi N}-0.070 g_7^{NN})\big]^2\nn\\
&\quad \times \left|y_{e\Psi} (y_{qu}+y_{qd})\lambda_{ed}\lambda_{u\Psi}\right|^2   \times \left(\dfrac{\text{TeV}^5}{m_S^2 m_R^2 m_\Psi}\right)^2~\text{year}^{-1}\;,
\end{align}
where all of the masses are normalized by the scale of $1\tev$.

The most stringent $\onbb$-decay limit comes from the experiment KamLAND-Zen~\cite{KamLAND-Zen:2016pfg}:  $T_{1/2}^{0\nu}>1.07\times 10^{26}~\text{year}$ at 90\% confidence level (C.L.). The constraint from the final results of GERDA experiment~\cite{GERDA:2020xhi} is slightly weaker. 
There exist a number of planned experiments at the ton-scale~\cite{Kharusi:2018eqi,Abgrall:2017syy,CUPIDInterestGroup:2019inu,Paton:2019kgy,Chen:2016qcd,Adams:2020cye} 
that aim to improve the half-life sensitivity  by about 2 orders of magnitude, reaching  $T_{1/2}^{0\nu}> 10^{28}~\text{year}$ at 90\% C.L.
From Eq.~\eqref{eq:half-life_2}, 
we thus infer that future ton-scale $\onbb$-decay experiments are able to probe TeV-scale LNV with the masses $m_S=m_R=m_{\Psi}=1\tev$ and the couplings $|y_{e\Psi}(y_{qu}+y_{qd})\lambda_{ed}\lambda_{u\Psi}|\gtrsim (3-9)\times 10^{-3}$ for $g_V^{\pi N}=\tilde g_V^{\pi N} =1$ and $g^{NN}_6=g^{NN}_7=\pm 1/2$.
In Fig.~\ref{fig:DBD-sensitivity}, we show for three sets of assumed values for these low-energy constants, the half-lives of $\onbb$ decay for $^{136}$Xe as a function of the LNV scale. Clearly, we can see that typically the LNV scale up to around $4.5-6\tev$ for the Wilson coefficients $C_{Qu1}+C_{Qd1}=1$ is in the reach of next-generation ton-scale $\onbb$-decay experiments. 
The same figure also illustrates that the sensitivity of the inferred LNV scale to these LECs is sizable.

\begin{figure}[!htb] 
\centering
\includegraphics[width=0.45\textwidth]{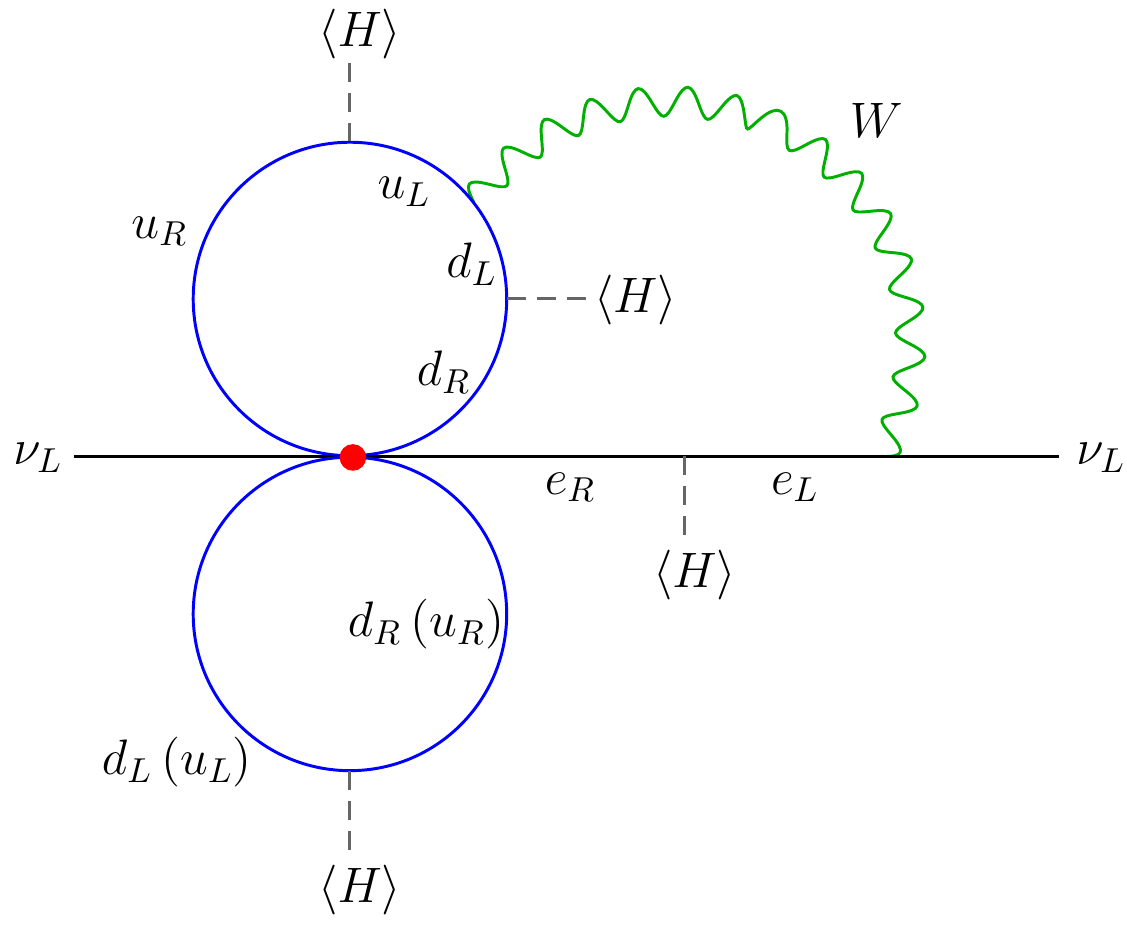}	
\caption{The LNV vector operators contribute to neutrino masses at three-loop or higher order, and are completely 
negligible. Shown here is a  typical three-loop Feynman diagram contributing to the neutrino mass. The red dot depicts the dimension-9 operators in Eq.~\eqref{eq:dimension-9_neutrino}. The analogous diagram in the simplified model is shown in 
Fig. \ref{fig:neutrino_mass_full} of Appendix \ref{app:neutrino_mass}.}
\label{fig:neutrino_mass_dim9}
\end{figure}

\subsection{Neutrino masses: dimension-9 operators}
\label{sec:neutrino-masses-dim9}
Before closing this section, it is interesting to discuss the connection in this model between $\onbb$ decay and neutrino masses. On general grounds, the Schechter-Valle theorem~\cite{Schechter:1981bd} implies that the observation of $\onbb$ decay implies the existence of a light neutrino Majorana mass term. For the model of interest here, the $\onbb$-decay operators can induce Majorana neutrino masses at loop level.
It is instructive to consider this connection after integrating out the heavy degrees of freedom. Retaining only the light degrees of freedom, we obtain the loop-induced Majorana mass 
from diagrams such as Fig.~\ref{fig:neutrino_mass_dim9}.
In the simplified model neutrino masses are generated at three-loop level \footnote{It is evident from Fig. \ref{fig:neutrino_mass_full} that if $S$ acquires a vacuum expectation value (vev), 
then a neutrino mass is generated at two-loop order. With $\Lambda_{\text{UV}} \sim \Lambda$, the estimate for the neutrino 
mass from such a two-loop diagram is larger by a factor of $16 \pi^2 \langle S \rangle/m_d$ compared to the estimate from the three-loop diagram. 
In this case the neutrino mass is still 
suppressed far below its actual value. 
If a vev for $S$ doesn't occur at tree-level, at one-loop a term 
$S^\dagger H$ in the Lagrangian is generated, which leads to $\langle S \rangle \sim m_d \Lambda^2_{\text{UV}}/(16 \pi^2 m^2_S)$. With $\langle S \rangle$ of this size, then the direct three-loop diagram (\ref{fig:neutrino_mass_full}) and two-loop $\langle S \rangle$ induced neutrino masses are parametrically the same size.
} as briefly discussed in Appendix~\ref{app:neutrino_mass}.  The red dot in Fig.~\ref{fig:neutrino_mass_dim9} depicts the effective operators in the form of 
\begin{align}
\label{eq:dimension-9_neutrino}
&\frac{C_{Qu1}}{\Lambda^5}(\bar{ u}_R u_L) (\bar{u}_R\gamma_\mu d_R)(\bar{e}_R \gamma^\mu \nu_L^{c})\;,\nn\\
&\frac{C_{Qd1}}{\Lambda^5}(\bar{ d}_L d_R) (\bar{u}_R\gamma_\mu d_R)(\bar{e}_R \gamma^\mu \nu_L^{c})\;.
\end{align}
In the SMEFT, 
they arise from the $\nu-$component of the $SU(2)_L \times U(1)_Y$ invariant dimension-9 effective operators in Eq.~\eqref{eq:dimension-9_SMEFT}.  The contribution of these operators to the neutrino mass is highly 
suppressed by three insertions of the light quark masses and one of the electron mass, as well 
as by the loop factors, and is estimated as 
\begin{eqnarray}
    m_\nu &\sim& \frac{m_d m_u m_e}{(16 \pi^2)^3} \frac{\Lambda_\text{UV}^2}{\Lambda^5}
    \left(m_u C_{Qu1} + m_d C_{Qu2}\right) \nn \\
    & \simeq &10^{-24} \text{MeV} \left(\frac{m^3_{u,d} m_e}{\text{MeV}^4}\right)\left(\frac{\Lambda_\text{UV}}{\text{1 TeV}}\right)^2 
    \left(\frac{\text{1 TeV}}{\Lambda}\right)^5\;,
\end{eqnarray}
which is completely negligible compared to the actual neutrino masses. 
Here $\Lambda_\text{UV}$ is an UV cutoff, with $\Lambda_\text{UV} \sim \Lambda$. Two powers 
of $\Lambda_\text{UV}$ arise from the quadratic divergence in the ``bubble'' loop of light quarks appearing below the ``horizontal line'' in Fig. \ref{fig:neutrino_mass_dim9}. (In dimensional regularization two powers of the light quark masses would appear instead.) In obtaining this estimate we have defined 
$\Lambda^5  \equiv (m^2_S m^2_R m_\Psi)$ and set equal to one the product of 
Yukawa couplings appearing in the Wilson 
coefficients $C_{Qu1}$ and $C_{Qd1}$: 
$\frac{1}{2} y_{e\Psi}\lambda_{ed}\lambda_{u\Psi}y_{qu} \equiv \frac{1}{2} y_{e\Psi}\lambda_{ed}\lambda_{u\Psi}y_{qd}
\equiv 1.$
The simplified model presented here 
cannot generate the observed neutrino masses, which therefore must arise from some other source.

We end by returning to the Schechter-Valle ``black box" theorem~\cite{Schechter:1981bd}.
It is noted that the contribution to Majorana neutrino masses 
arising from an insertion of the $\onbb$ decay ``black-box operator'' occurs at four-loop level.
The induced shift in the Majorana 
neutrino mass,
derived using the experimental limit $T^{0 \nu}_{1/2}>1.07\times 10^{26}$ year, is roughly $2\times 10^{-28}$~eV~\cite{Duerr:2011zd}. Here we have updated the numerical results of 
Ref. \cite{Duerr:2011zd} to reflect the more recent KamLAND-Zen limit \cite{KamLAND-Zen:2016pfg}.
This value is much smaller than the direct contribution to neutrino masses arising
in the 
simplified model considered here.

\subsection{Dimension-7 operator}
\label{sec:model_DBD-dim7}

Here we consider the effects of the interaction 
\begin{align}
    y_{e \Psi}^\prime \bar{\Psi}_L H e_R ~
\end{align}
on $\onbb$ decay. This interaction appears in the last line of the Lagrangian in Eq.~\eqref{eq:interactions}, and is allowed by all symmetries. It does not contribute to the LNV dimension-9 operators, but its presence 
does generate a LNV dimension-7 operator above the electroweak scale. The Yukawa coupling $y_{e \Psi}^\prime$ has zero lepton number. 

To see that it cannot be forbidden by any symmetry, note that the model already has Yukawa couplings of $S$ to the SM quarks. 
Under any discrete symmetry $S$ and $H$ must have the same charge, which means we cannot distinguish $S$ from $H$. 
Thus if we have $\bar{\Psi}_L S e_R$ in the Lagrangian we cannot forbid $\bar{\Psi}_L H e_R$. 
This operator cannot be forbidden by any $Z_N$ discrete symmetry. 

\begin{figure}[!htb] 
\centering
\includegraphics[width=0.4\textwidth]{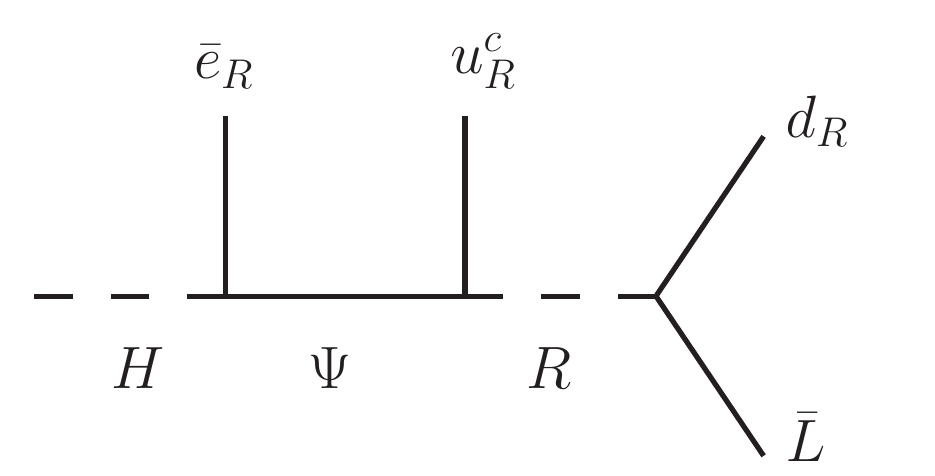}	
\caption{Quark-level Feynman diagram that induce $\Delta L=2$ dimension-7 $\onbb$-decay operator $O_{Leu\bar{d}H}$ at low energies. 
}
\label{fig:feyn_DBD_quark-dim7}
\end{figure}

Integrating out $S$ and $R$ at tree-level gives the following new contribution to the effective theory below 
that mass scale,
\begin{align}
\label{eq:SMEFT-7}
\mathcal{L}_{{\rm SMEFT}}^{(7)} = & \dfrac{1}{\Lambda^3} C_{Leu\bar{d} H} O_{Leu\bar{d}H} + \text{h.c.} \;,
\end{align}
where the dimension-7 operator is given by 
\begin{align}
    O_{Leu\bar{d}H} = (\bar{u}_R\gamma_\mu d_R)(\bar{e}_R \gamma^\mu L^{c}) H^*
\end{align}
with a Wilson coefficient 
\begin{align}
\label{eq:Wilson-coefficient-dim-7}
\dfrac{C_{Leu\bar{d} H}}{\Lambda^3}= \dfrac{1}{2} \dfrac{y_{e\Psi}^\prime\lambda_{ed}\lambda_{u\Psi}}{m_R^2 m_\Psi}
\end{align}
The Feynman diagram generating this effective interaction is shown in Fig. \ref{fig:feyn_DBD_quark-dim7}. 
Note that this Wilson coefficient depends on the same combination of Yukawa couplings $(\lambda_{ed}\lambda_{u\Psi})$
that appears in the Wilson coefficients of the dimension-9 operators. Of course, here $y_{e\Psi}^\prime$
also occurs, which does not appear in the dimension-9 Wilson coefficients.

\begin{figure}[!htb] 
\centering
\includegraphics[width=0.4\textwidth]{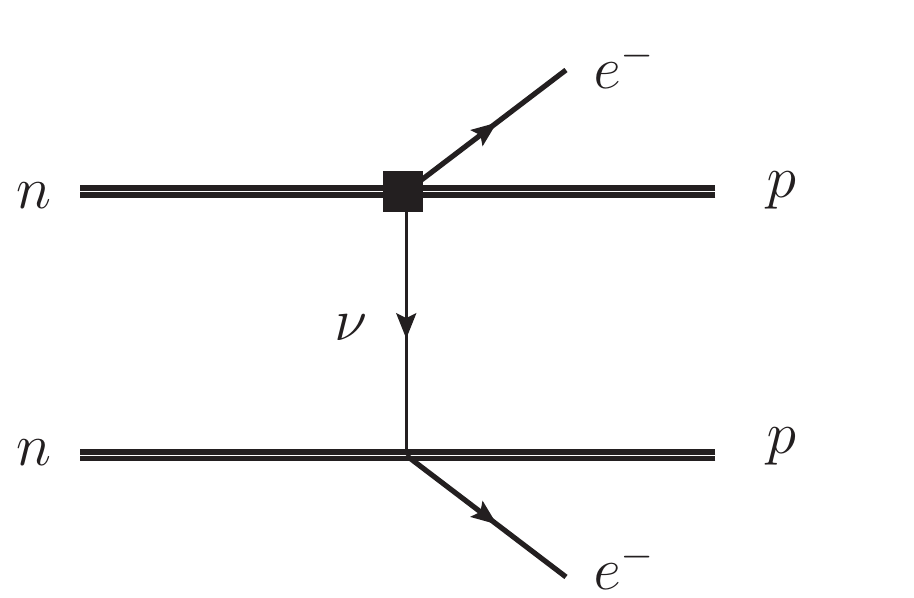}	
\caption{Hadron-level Feynman diagram 
for $\onbb$ decay, 
induced by the LNV operator $np \overline{e} \nu^c$ denoted by the black square. The other 
vertex denotes the single $\beta$ decay operator from SM interactions.
}
\label{fig:feyn_DBD_hadron-dim7}
\end{figure}

The operator $O_{Leu\bar{d}H}$ is proportional to a current of quarks, so the QCD RG evolution of this operator is trivial. 
After the electroweak symmetry breaking, the Higgs doublet $H$ develops the vev $\langle H \rangle = v/\sqrt{2}$, and the following effective Lagrangian in the LEFT 
is obtained
\begin{align}
\label{eq:SMEFT-7-EW}
\mathcal{L}_{{\rm LEFT}}^{(6)} = \dfrac{1}{v^2}& C^{(6)}_{\text{VR}}  O^{(6)}_{\text{VR}}
+ \text{h.c.} \;,
\end{align}
where the dimension-6 operator is 
\begin{align}
O^{(6)}_{\text{VR}} = (\bar{u}_R\gamma_\mu d_R)(\bar{e}_R \gamma^\mu  \nu^{c})\;.
\end{align}
and the Wilson coefficient is given by
\begin{align}
\label{eq:Wilson-CVR6}
\dfrac{1}{v^3} C^{(6)}_{\text{VR}} = & \dfrac{1}{\sqrt{2} \Lambda^3}  C^*_{Leu\bar{d} H}\;.
\end{align}
Due to the presence of the neutrino, this interaction make a long-distance contribution to 
$\onbb$ decay as shown in Fig. \ref{fig:feyn_DBD_hadron-dim7}. 

Below the scalar $\Lambda_\chi \sim 1\gev$, the quark-lepton operator are mapped onto hadron-lepton operators. Up to NLO in chiral expansion, the resulting one-body current for single $\beta$ decay was shown explicitly in Ref.~\cite{Cirigliano:2017djv}.

At LO in the Weinberg power counting, the dimension-6 operator $O^{(6)}_{\text{VR}}$ makes a vanishing contribution to  $\onbb$ decay.
The first non-vanishing contribution is obtained by expanding the one-body vector 
and axial currents to NLO \cite{Cirigliano:2017djv}. The contribution of the dimension-6 operator $O^{(6)}_{\text{VR}}$ to the amplitude for $0^+\to 0^+$ is given by~\cite{Cirigliano:2017djv, Cirigliano:2018yza}
\begin{align}
   \mathcal{A}_{\text{VR}}= \dfrac{g_A^2 G_F^2 m_e }{\pi R_A}\left[ \mathcal{A}_E \bar{u}(k_1) \gamma_0 C \bar{u}^T (k_2) \dfrac{E_1-E_2}{m_e}+ \mathcal{A}_{m_e} \bar{u}(k_1) C \bar{u}^T (k_2) \right] 
\label{eq:amplitude-dim7-operator}\;,
\end{align}
where $E_1$ and $E_2$ are the energies of the emitted electrons and
the reduced amplitudes  are
\begin{align}
    \mathcal{A}_E = C^{(6)}_{\text{VR}} M_{E,R}\; , \quad \mathcal{A}_{m_e} = C^{(6)}_{\text{VR}} M_{m_e,R}\;,
\end{align}
and the NMEs 
\begin{eqnarray}
\label{eq:NME-dim7}
M_{E,R} &=& -\dfrac{1}{3} \left[\dfrac{g^2_V}{g^2_A} M_F - \dfrac{1}{3}\left(2M^{AA}_{GT}+M^{AA}_{T}\right)  \right]\;, \nn\\
M_{m_e,R} &=& \dfrac{1}{6} \left[\dfrac{g^2_V}{g^2_A} M_F +\dfrac{1}{3}\left(M^{AA}_{GT}-4M^{AA}_{T}\right) 
+ 3 ( M^{AP}_{GT}+M^{PP}_{GT} + M^{AP}_T + M^{PP}_T)\right] \;.
\end{eqnarray}
Here, $g_V=1$, and the six long-range NMEs are given by $M_F=-0.89$, $M^{AA}_{GT}=3.16$, $M^{AP}_{GT}=-1.19$, $M^{PP}_{GT}=0.39$, 
$M^{AP}_T=-0.28$, and $M^{PP}_T=0.09$ for $^{136}$Xe calculated using the QRPA method \cite{Hyvarinen:2015bda}. 
The NME $M^{AA}_T$ cannot be extracted from current calculations of light and heavy neutrino Majorana 
mass contributions to $\onbb$ decays of current experimental interest. Here we set it to zero as it is expected to be small 
compared to the other NMEs. \footnote{In the case of light nuclei, 
Variational Monte Carlo (VMC) 
calculations for $-M^{AA}_T \simeq 0.06-0.25$ are obtained in \cite{Pastore:2017ofx}. Among
$\Delta I=2$ nuclear isospin transitions, the largest value found is $-M^{AA}_T \simeq 0.15$.}
With these as input, $M_{E,R}=0.89$ and $M_{m_e,R}=-0.41$, again 
for $^{136}$Xe.

The amplitude is proportional to the outgoing electron energies or the electron mass and therefore 
highly suppressed. In the Weinberg power counting three-body nucleon interactions and loops of pions are expected to make comparable 
contributions. As a result, the limit and projection obtained below should be viewed as an 
order-of-magnitude estimate \cite{Cirigliano:2017djv}.

The inverse half-life is given by~\cite{Cirigliano:2017djv}
\begin{align}
\label{eq:half-life_dim7}
\left(T_{1/2}^{0\nu}\right)^{-1}= g_A^4 \left[ 4 G_{02} |\mathcal{A}_E |^2 
- 4G_{03} \text{Re} \left( \mathcal{A}_{m_e} \mathcal{A}_E \right) + 2 G_{04} |\mathcal{A}_{m_e} |^2 \right] 
\end{align}
with the phase space factors $G_{02}=3.2\times 10^{-14}$ year$^{-1}$, $G_{03}=0.86\times 10^{-14}$ year$^{-1}$, 
and $G_{04}=1.2\times 10^{-14}$ year$^{-1}$ for $^{136}$Xe. Substituting for $C^{(6)}_{\text{VR}}$, the NMEs, and the
phase space factors, gives 
\begin{eqnarray}
 \left(T_{1/2}^{0\nu}\right)^{-1}
 &=&8.6 \times 10^{-18} |y_{e\Psi}^\prime \lambda_{ed} \lambda_{u\Psi}|^2 
  \times \left(\dfrac{\text{TeV}^3}{m_R^2 m_\Psi}\right)^2~\text{year}^{-1}\;.
\end{eqnarray}

\begin{figure}[!htb] 
\centering
\includegraphics[width=0.6\textwidth]{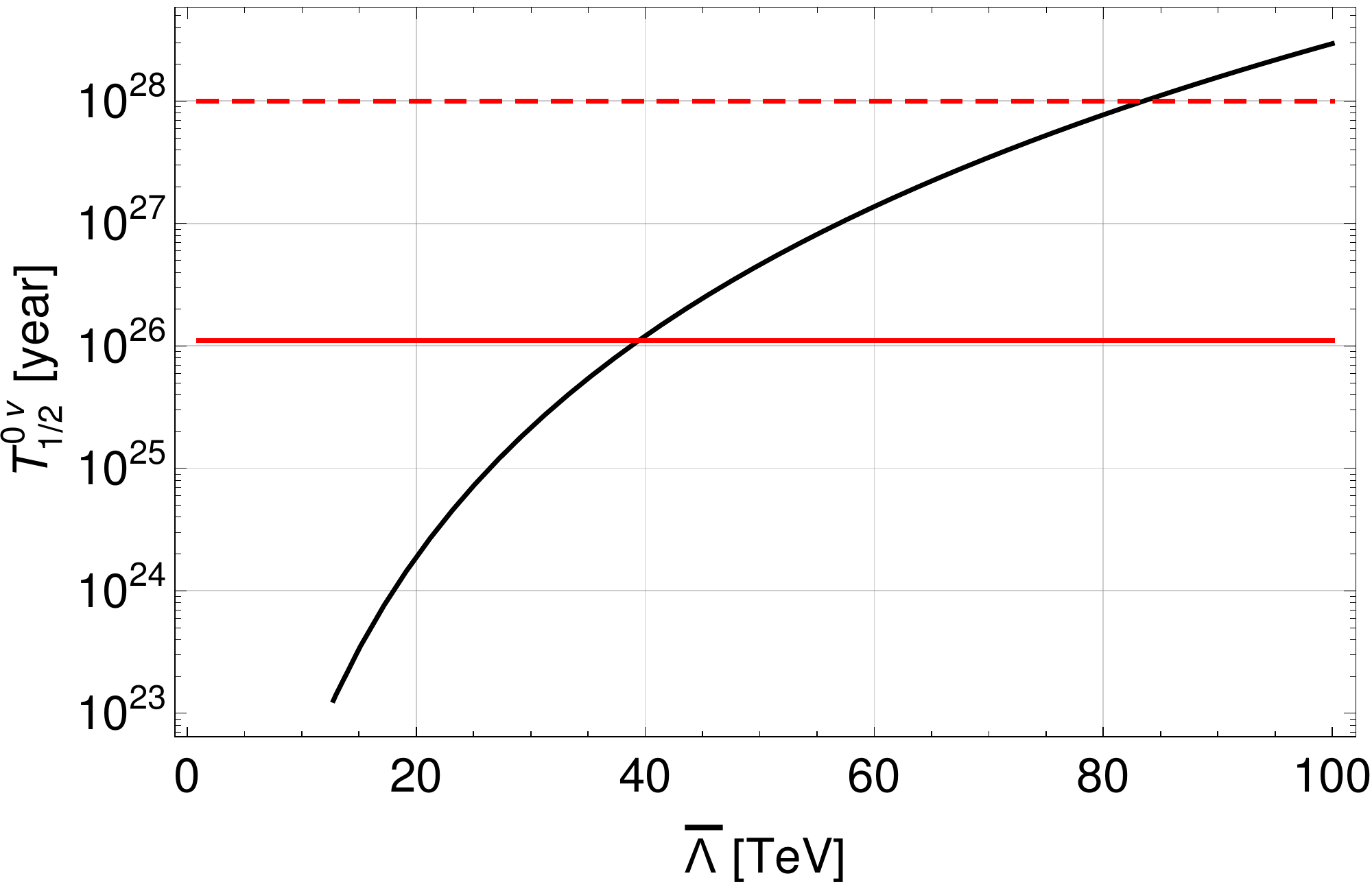}	
\caption{ 
Half-lives of $\onbb$ decay as a function of the effective LNV scale $\overline\Lambda\equiv \Lambda/\left|C_{Leu\bar{d}H}\right|^{1/3}$.
 The solid red and dashed red lines correspond to the constraints on the half-life in the KamLAND-Zen and future ton-scale experiments for $^{136}$Xe. 
}
\label{fig:C7-limits}
\end{figure}

The Kamland-Zen limit of $T_{1/2}^{0\nu}> 1.07 \times 10^{26}$ year gives a constraint of 
$| y_{e\Psi}^\prime \lambda_{ed} \lambda_{u\Psi}| < 3.3\times 10^{-5}$ for $m_R=m_\Psi=1$ TeV. 
Setting the Yukawa couplings equal to one, this bound translates into a bound on 
$\Lambda \equiv (m^2_R m_\Psi)^{1/3} \gsim 40$ TeV.
Current and projected limits 
are shown in Fig. \ref{fig:C7-limits}, where we can see that future ton-scale experiments are sensitive to the LNV scale up to $90\tev$ for the Wilson coefficient $C_{Leu\bar{d}H}=1$.

One can see that $\onbb$ decay is more sensitive to the dimension-7 operator  $O_{Leu\bar{d}H}$ compared to the dimension-9 vector operators. Following Refs.~\cite{Cirigliano:2017djv,Cirigliano:2018yza}, this can be understood in chiral power counting. Denoting $\epsilon_\chi = q/\Lambda_\chi$ with $q\sim m_\pi$ the typical momentum transfer of $\onbb$ decay, 
then the typical outgoing electron energies scale as $E \sim \hbox{MeV} \sim \Lambda_\chi \epsilon_\chi^3$. Since
the reduced amplitudes $m_e \mathcal{A}_E$ and $m_e \mathcal{A}_{m_e}$ induced by $O_{\text{VR}}^{(6)}$ are
proportional to the electron energies, they scale as
$\Lambda_\chi \epsilon_\chi^3$. In contrast,
$m_e \mathcal{A}_M$ induced by the dimension-9 vector operators 
scales as $\Lambda_\chi^2 \epsilon_\chi^2/v $. Although all of them are chirally suppressed, $\mathcal{A}_M$ is smaller than $\mathcal{A}_E$ and $ \mathcal{A}_{m_e}$ by a factor of $\Lambda_\chi /(v\epsilon_\chi)\simeq 0.03$.

The existence of the  Yukawa interaction $y_{e\Psi}^\prime \bar{\Psi}_L H e_R$ would also contribute to the deviation from unitarity of the Pontecorvo-Maki-Nakagawa-Sakata matrix and mass mixing of charged leptons with more phenomenological implications and constraints. However, since the LHC searches considered in the next section are insensitive to this interaction, 
we simply assume that it is suppressed, which might be realized in UV theories,  and focus on the interplay of chirally suppressed $\onbb$ decay induced by the dimension-9 vector operators and LHC searches.

\subsection{Neutrino masses: dimension-7 operators}
\label{sec:neutrino-masses-dim7}
This discussion closes with some comments about neutrino masses. Like the dimension-9 operator previously discussed in Section 
\ref{sec:model_DBD-dim9}, 
the dimension-7 operator Eq. (\ref{eq:SMEFT-7}) also
generates neutrino masses at higher loop order. In particular, a neutrino mass is first generated 
at two-loops and is finite. It is highly suppressed due to one electron mass and two light quark mass 
insertions, and one estimates that 
\begin{eqnarray}
    m_\nu &\sim& \frac{C_{Leu\bar{d}}}{(16 \pi^2)^2}
    \frac{m_d m_u m_e}{\Lambda^3} 
    v \nn \\
    & \simeq &10^{-17}  C_{Leu\bar{d}} \text{MeV} \left(\frac{m_{u} m_d m_e}{\text{MeV}^3}\right)
    \left(\frac{\text{1 TeV}}{\Lambda}\right)^3\;,
\end{eqnarray}
which is completely negligible. Consequently, the observed values of the neutrino masses do not 
constrain the size of this Wilson coefficient.

\section{LHC searches}
\label{sec:LHC}

We will investigate the complementary test of TeV-scale LNV in our simplified model using LHC searches.  The process with a pair of first generation leptons with the same charge ($e^\pm e^\pm$) and at least two jets would provide a clear sign of LNV at the LHC.
Searches for this same-sign dilepton plus dijet signal have thus far yielded null results, leading to constraints on different models for TeV-scale LNV. We analyze the corresponding implications for our simplified model below. Additionally, searches for leptoquarks as well as those for dijet resonances, which do not rely on LNV signatures,
can also be used to extend the sensitivities to the masses and couplings of new particles in our simplified model. 

In this section, we will reinterpret the existing searches, which are performed at the 13~TeV LHC with integrated luminosities of $\sim 40\fbi-140\fbi$, in terms of the corresponding LNV and lepton number conserving processes generated in our model, then make projections for the high-luminosity LHC (HL-LHC) 
with the integrated luminosity of $3\abi$. 
In all of our projections for exclusion and discovery at the HL-LHC, we make the 
strong assumption
that the selection cuts and efficiencies remain unchanged from those in existing searches. The reaches presented here might be improved with an optimization of cuts.

Note that both $y_{qu}$ and $y_{qd}$ terms in Eq.~\eqref{eq:interactions} can contribute to the scalar production at the LHC, which do not interfere. For the charged scalar $S^\pm$, their contributions to the production cross section are the same if $y_{qu}=y_{qd}$. For the neutral scalar $S^0$, however, the contribution to the production cross section from the $y_{qu}$ term is larger than that from the $y_{qd}$ term, since $u$-quark parton distribution function (PDF) is 
typically about two times larger than $d$-quark PDF for the Bjorken $x\sim m_S/\sqrt{s}$, which is around $0.1-0.2$. Here, $\sqrt{s}=13\tev$ is the center-of-mass energy of the LHC.
Consequently, the constraints from dijet search in case of non-vanishing $y_{qd}$ is relatively weaker than that for non-vanishing $y_{qu}$.

Besides, different from the $\onbb$-decay half-life that depends on $(y_{qu}+y_{qd})^2$, there is no interference between the contributions from the $y_{qu}$ and $y_{qd}$ terms to the production cross sections of $pp \to S^\pm,S^0$. 
Hereafter, to simplify the presentation of our analysis, we assume $y_{qu}=0$ and $y_{qd}\neq 0$, and study the interplay of LHC searches and $\onbb$ decay. 
Our conclusions would be 
qualitatively similar if $y_{qd}=0$ and $y_{qu}\neq 0$ were assumed instead.

\subsection{Leptoquark searches}
\label{sec:LQ_search}

Direct search for pairs of first-generation leptoquarks  by the ATLAS collaboration
at the 13 TeV LHC with an integrated luminosity of $139\fbi$ has excluded the leptoquark mass region below 1.8~TeV~\cite{Aad:2020iuy}.
To make a conservative projection for the future sensitivity to leptoquarks, 
we assume that both observed upper limit on the number of signal events and the number of the background events in Ref.~\cite{Aad:2020iuy} scale with the integrated luminosity, and obtain that the lower limit on the leptoquark mass at the LHC is about $2\tev$ with an integrated luminosity of $1\abi$.  
These constraints become weaker if the decay branching ratio of leptoquark into electron $e^+/e^-$ and quark is less than 1.

In the analysis of the leptoquark search~\cite{Aad:2020iuy}, a pair of opposite-sign electrons $e^+e^-$ and at least two jets are required in the final state. 
In addition, the pairs of electron and jet closest in the invariant mass must satisfy $m^{\text{asym}}_{ej}\equiv (m_{ej}^{\text{max}}-m_{ej}^{\text{min}})/(m_{ej}^{\text{max}}+m_{ej}^{\text{min}})  <0.2$, where $m_{ej}^{\text{max}}$ and $m_{ej}^{\text{min}}$ denote the larger and smaller of the two electron-jet invariant masses. The upper limits on $\sigma\times \mathcal{B}\times A$
for each leptoquark mass are derived by performing fits to the distribution of $m^{\text{mean}}_{ej}$, which is defined as $m^{\text{mean}}_{ej}\equiv (m_{ej}^{\text{max}}+m_{ej}^{\text{min}})/2$. Here, $\sigma$ denotes the leptoquark pair production cross section, $\mathcal{B}$ is the product of the decay branching ratios of intermediate particles,
and $A$ describes the overall acceptance.

\begin{figure}[!htb] 
\centering
\includegraphics[width=0.7\textwidth]{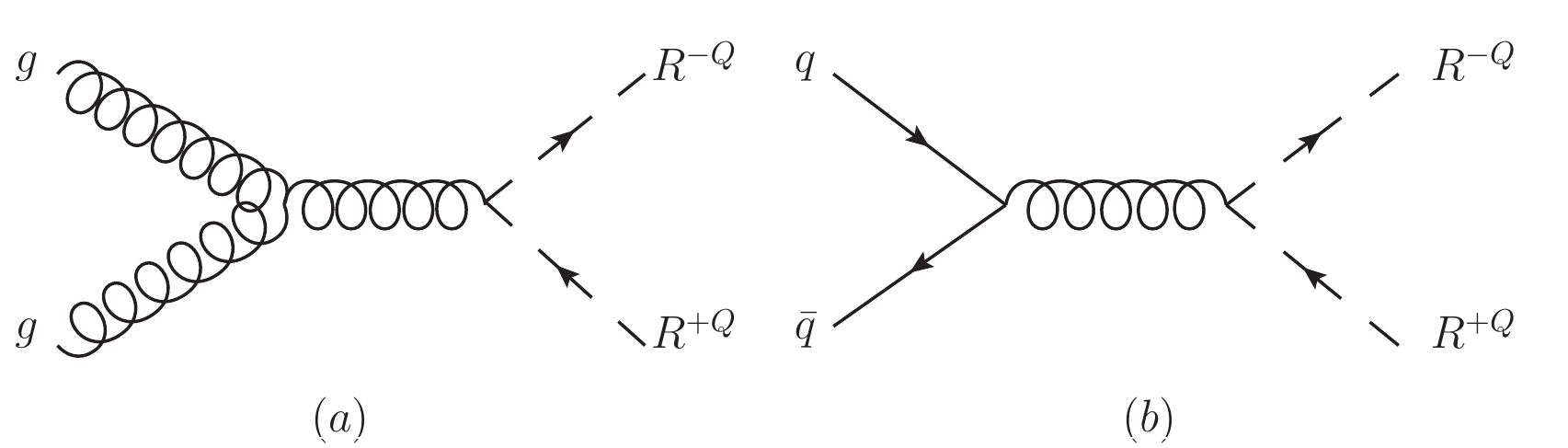}	
\caption{Representative Feynman diagrams for the parton-level pair production of leptoquarks $(R^{\pm Q})$.
Here, $g$ denotes a gluon $q$ denotes a quark, and the electric charge $Q=2/3,1/3$. 
}
\label{fig:LQ_production}
\end{figure}

\begin{figure}[!htb] 
\centering
\includegraphics[width=0.45\textwidth]{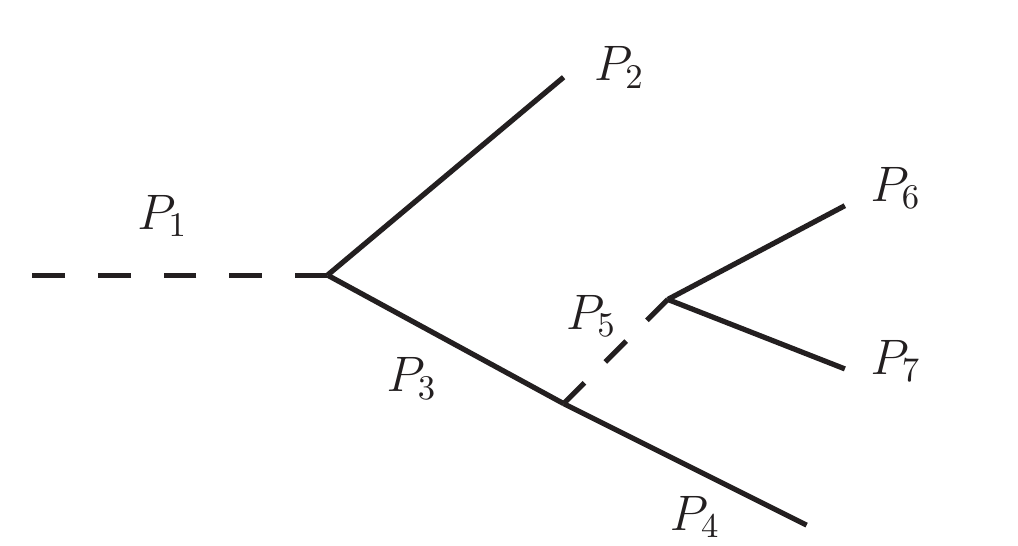}	
\caption{Feynman diagram for the cascade decays of leptoquarks. The labels $P_1,\ldots,P_7$ denote the possible particles in the chain, which are specified in Tab.~\ref{tab:lq_cascade}. 
}
\label{fig:LQ_cascade}
\end{figure}

\begin{table}[ht]
\caption{The cascade decays of leptoquarks in the presence of $\Psi$. }
\begin{center}
\begin{tabular}{c|c|c|c|c|c|c}
\hline 
\hline
$P_1$ & $P_2$ & $P_3$ & $P_4$ & $P_5$ & $P_6$ & $P_7$\\ \hline
${R}^{2/3}$ & $u$ & $\Psi^0$ & $\bar{u}$ & ${R}^{2/3}$ & $d$ & $e^+$  \\ \hline
${R}^{2/3}$ & $u$ & $\Psi^0$ & $e^-$ & $S^+$ & $u$ & $\bar{d}$  \\ \hline
${R}^{1/3}$ & $\bar{u}$ & $\Psi^+$ & $e^+$ & $S^0$ & $d$ & $\bar{d}$  \\ \hline
\hline 
\end{tabular}
\end{center}
\label{tab:lq_cascade}
\end{table}

The signal process $pp\to {R}^{+2/3}{R}^{-2/3}$, ${R}^{2/3} \to e^+ d $, ${R}^{-2/3} \to e^- \bar d$ has the same acceptance $A$ as Ref.~\cite{Aad:2020iuy}. Thus the upper limit on $\sigma \times \mathcal{B}$ reported in Ref.~\cite{Aad:2020iuy} can be directly applied to this signal process.
However, this conclusion is not true for other signal processes due to different cut acceptances. To be more specific, in our simplified model the leptoquark ${R}$ can also have a cascade decay if $m_R > m_\Psi$. In Figs.~\ref{fig:LQ_production} and \ref{fig:LQ_cascade}, we show schematic Feynman diagrams for the pair production and cascade decays of leptoquarks. 

In fact, for the other leptoquark pair production processes such as $pp\to {R}^{+2/3}{R}^{-2/3}$, ${R}^{-2/3} \to e^- \bar d$, ${R}^{+2/3} \to \Psi^0 u,$ $\Psi^0 \to e^+ \bar{u} d$, and $pp\to {R}^{+1/3}{R}^{-1/3}$, ${R}^{+1/3} \to \Psi^+ \bar{u} $, ${R}^{-1/3} \to \Psi^- u$, $\Psi^\pm \to e^\pm d \bar{d}$, 
after imposing the cut $m^{\text{asym}}_{ej}<0.2$
the efficiencies 
and the values of $m^{\text{mean}}_{ej}$ are smaller, since there are more jets in the final state. This leads to a smaller overall acceptance $A$.
Finally, it is noted that in our model the processes $pp\to e^+ \Psi^0$, $\Psi^0 \to e^- u\bar{d}$ and $pp\to e^- \bar\Psi^0$, $\bar\Psi^0 \to e^+ d\bar{u}$ also contribute to our reinterpretion 
of the leptoquark search~\cite{Aad:2020iuy}, but their contributions are substantially reduced by  the selection cut on $m_{ej}^{\text{asym}}$.  

After taking into account all of the signal processes with the possible decays of $R$, and deducing 
the corresponding acceptance, the lower bound on the leptoquark mass is expected to be smaller than that in Ref.~\cite{Aad:2020iuy}. Instead of reanalyzing 
the fits for all of these signals, we will assume conservatively the leptoquark mass $m_R = 2\tev$.

The leptoquark pair production $pp\to {R}^{+1/3}{R}^{-1/3}$ can also give signatures of jets $+$ MET (monojet), jets $+$ $e^\pm$ $+$ MET, where the missing energy (MET) comes from neutrino(s) in $R^{+1/3}\to \bar{d}\nu$ and/or $R^{-1/3}\to d\bar{\nu}$ and $e^\pm$ come from the cascade decay of $R^{\pm 1/3}$. However, due to low trigger efficiencies of these signals the constraints are expected to be very weak, and will not be considered hereafter.

Finally, it is worth noting that other searches for first-generation leptoquark may give constraints comparable to the pair production with the assumption that both the leptoquark coupling $\lambda_{ed}$ and decay branching ratio are equal to 1. In Ref.~\cite{Khachatryan:2015qda}, a search for the single production of leptoquark was performed by the CMS Collaboration at the 8~TeV LHC with an integrated luminosity of $19.6\fbi$. The recast and projection of this search with the same selection cuts being imposed~\cite{Schmaltz:2018nls} show that  $m_R < 1.4\tev$ could be excluded at the 13~TeV LHC with an integrated luminosity of $36\fbi$. For comparison, the search for leptoquark pair production has excluded $m_R < 1.435\tev$ by the CMS Collaboration with an integrated luminosity of $35.9\fbi$~\cite{Sirunyan:2018btu}. Ref.~\cite{Shen:2022qp} finds that $m_R< 3.6\tev$ would be able to be excluded at the HL-LHC with an integrated luminosity of $3\abi$ with the boosted decision tree method being used in the analysis. It was found in Refs.~\cite{Schmaltz:2018nls,Crivellin:2021egp} that other constraints on $m_R$ can be obtained in the searches for single resonant production and Drell-Yan production, which could exclude  $m_R < 2.5\tev$ and $m_R <3.8\tev$  at the 13~TeV LHC with the integrated luminosities of $36\fbi$ and $139\fbi$, respectively, and $m_R < 4.2\tev$ with the integrated luminosity of $3\abi$ in the former process. 

One might thus ask if $m_R = 2\tev$ that we choose for our benchmarks satisfies the current constraints in these leptoquark searches. It is important to note that the cross sections of (1) the single production of leptoquark and (2) single resonant production of leptoquark are proportional to $|\lambda_{ed}|^2$, and the cross section of (3) the Drell-Yan production mediated by exchange of leptoquark is proportional to  $|\lambda_{ed}|^4$. By decreasing $\lambda_{ed}$, these constraints will be significantly released. With integrated luminosities of $36\fbi$, $36\fbi$ and $139\fbi$ at the 13~TeV LHC, $m_R < 0.9 \tev$, $1.8\tev$, $1.5\tev$ with $\lambda_{ed}=0.6$, $0.4$, $0.6$ for the processes (1) (2) (3), respectively~\cite{Crivellin:2021egp}.
On the other hand, the impact on the mass reach for the pair production of leptoquarks is much smaller.

\subsection{Same-sign dilepton plus dijet search}
\label{sec:SSDL}

\begin{figure}[!htb] 
\centering
\includegraphics[width=0.4\textwidth]{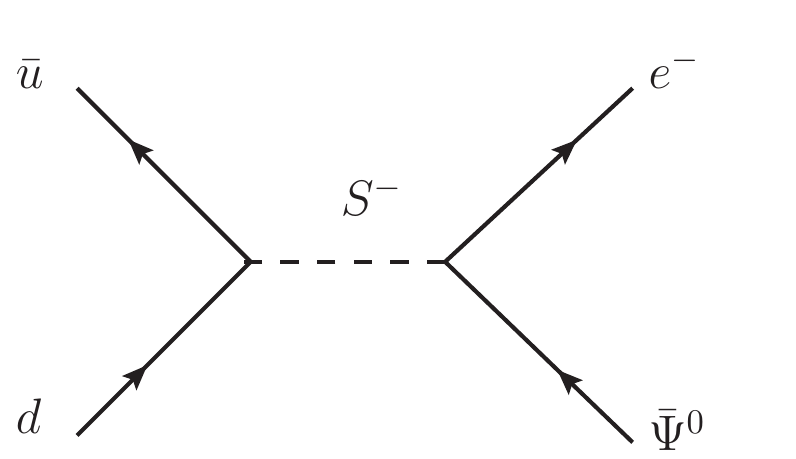}	
\caption{Feynman diagram for the parton-level process $\bar{u}d\to e^- \bar{\Psi}^0$ in SS-1. 
}
\label{fig:SSDL-Feyn}
\end{figure}

In order to directly test TeV-scale LNV at the LHC, we study the same-sign dilepton plus dijet (SSDL) processes. 
In our simplified model, lepton number is violated if $\lambda_{ed} \lambda_{u\Psi}\neq 0$, which translates into 
(1) if $\bar{\Psi}^0\to e^- \bar{d} u$ or its charge conjugate occurs since the lepton number of $\Psi^0$ is equal to 1; (2) if  ${R}^{-2/3} \to e^- \bar d$ and ${R}^{+2/3} \to \Psi^0 u$ or their charge conjugates occur simultaneously.
To this end, we will consider the following signal processes (with processes having charge-conjugate final states  not shown explicitly): 
\begin{itemize}
\item SS-1: $pp \to e^- \bar{\Psi}^0$, $\bar{\Psi}^0\to e^- \bar{d} u$. It proceeds 
through an $s$-channel production of $S^-$, followed by a LNV (cascade or three-body) decay of $\bar{\Psi}^0$. 
\item SS-2: $pp\to {R}^{+2/3}{R}^{-2/3}$, ${R}^{-2/3} \to e^- \bar d$, ${R}^{+2/3} \to \Psi^0 u,$ $\Psi^0 \to e^- \bar{d} u$ if $m_R > m_\Psi$. Here the decay of $\Psi^0$ is lepton number conserving. 
\item SS-3: $pp\to {R}^{+2/3}{R}^{-2/3}$, ${R}^{-2/3} \to \bar{\Psi}^0 \bar{u}$, ${R}^{+2/3} \to \Psi^0 u$, $\Psi^0 \to e^- \bar{d} u$ and $\bar{\Psi}^0 \to e^- \bar{d} u$ if $m_R > m_\Psi$. 
Here the decay of $\Psi^0$ is lepton number conserving 
and the decay of $\bar{\Psi}^0$ is lepton number violating.
\end{itemize}
The parton-level processes $\bar{u}d\to e^-\bar{\Psi}^0$ in SS-1 and $gg,q\bar{q}\to R^{+2/3}R^{-2/3}$ in SS-2 and SS-3 are shown in Fig.~\ref{fig:SSDL-Feyn} and Fig.~\ref{fig:LQ_production}, respectively. The decays of $\bar{\Psi}^0$, $R^{+2/3}$ and their anti-particles can be read off in Fig.~\ref{fig:LQ_cascade} and Table~\ref{tab:lq_cascade}. 

Searches for the right-handed $W$ boson and heavy neutrino at the 13~TeV LHC with the integrated luminosities of  $36.1\fbi$~\cite{ATLAS:2018dcj} and $35.9\fbi$~\cite{CMS:2018agk} have been performed by the ATLAS and CMS Collaborations, respectively. The ATLAS analysis~\cite{ATLAS:2018dcj} is divided into the same-sign and opposite-sign channels with $e^\pm e^\pm$ and $e^+ e^-$, respectively. 
In comparison with the SSDL search, the opposite-sign dilepton search~\cite{ATLAS:2018dcj}  suffers from much larger SM backgrounds, and is always less sensitive to the couplings $y_{e\Psi}$ and $y_{qd}$ unless the magnitudes of the couplings satisfy $|\lambda_{ed,u\Psi}|\ll |y_{e\Psi,qd}|$. In the CMS analysis~\cite{CMS:2018agk} there is no charge requirement on electrons, so that its sensitivity cannot compete with that of the ATLAS analysis.

\begin{table}[ht]
\caption{The benchmark points that we choose for the collider and $\onbb$-decay studies.}
\begin{center}
\begin{tabular}{c|c|c|c|c}
\hline 
\hline
    & $m_\Psi$ & $m_S$ & $m_R$ & $\sigma_{S^\pm}$  \\ \hline
BP1 &  $1.0\tev$ & $2.0\tev$ & $2.0\tev$ & $3.2y_{qd}^2~\text{pb}$ \\ \hline
BP2 &  $1.9\tev$ & $2.0\tev$ & $2.0\tev$ & $3.2y_{qd}^2~\text{pb}$ \\ \hline
BP3 &  $1.0\tev$ & $4.5\tev$ & $2.0\tev$ & $6.4y_{qd}^2~\text{fb}$ \\ \hline
BP4 &  $3.0\tev$ & $3.5\tev$ & $2.0\tev$ & $78y_{qd}^2~\text{fb}$ \\ \hline
\hline 
\end{tabular}
\end{center}
\label{tab:Benchmark}
\end{table}

In Tab.~\ref{tab:Benchmark}, we show four benchmark points for $m_R=2\tev$ and different $m_S$ and $m_\Psi$ assuming $m_S > m_\Psi$ to ensure that the  cross section of the SSDL signal is sizable enough. The cross section of $pp\to {R}^{+2/3}{R}^{-2/3} $ is $ 0.0155\fb$ at the 13~TeV LHC for the leptoquark mass $m_R=2\tev$~\cite{Dorsner:2016wpm}, while that of $pp\to S^\pm$ depends on the value of $y_{qd}$, as shown in the last column. After taking into account the decay branching ratios, we find that SS-2 and SS-3 always give negligible contributions compared to SS-1 unless the magnitudes of the couplings satisfy $|\lambda_{ed,u\Psi}|\ll |y_{e\Psi,qd}|$ or  $|\lambda_{ed,u\Psi}|\sim  |y_{e\Psi,qd}|  \ll 1$.

If $|\lambda_{ed,u\Psi}|\ll |y_{e\Psi,qd}|$, 
the branching ratio of the lepton number violating decay $\bar \Psi^0 \to e^- \bar{d} u$ is proportional to $ \lambda^4/y^4 \ll 1$, where $\lambda\sim \lambda_{ed,u\Psi}$ and $y\sim y_{qd,e\Psi}$ (for more details, the reader is referred to 
Appendix \ref{app:prod_decay}),  so that the cross sections of SS-1 and SS-3 are suppressed by the decay branching ratios compared to SS-2.
If $|\lambda_{ed,u\Psi}|\sim  |y_{e\Psi,qd}|  \ll 1$, the cross section of SS-1 is suppressed by small $|y_{qd}|^2$, while those of SS-2 and SS-3 are not suppressed.
In either case, the contribution of SS-2 and SS-3 to the SSDL signal can be comparable to that of SS-1. Nevertheless, in these two limits the sensitivity of the SSDL signal is very suppressed due to the small signal cross section,
and we have checked that for the 
benchmark points in Tab.~\ref{tab:Benchmark} there is 
no interplay of the SSDL search and $\onbb$ decay \footnote{For illustration, we assume $y_{qd}=y_{e\Psi}/4=0.1$, which satisfies the existing dijet constraint as we shall see, and set  $\lambda_{ed}=\lambda_{u\Psi}\equiv \lambda$. For BP2,
the cross section of SS-2 is comparable to that of SS-1 if $\lambda=0.025$.
Compared to the case of $\lambda=1$, however, the total cross section of the SSDL signal is smaller by an order of $10^3$ and the $\onbb$-decay rate is suppressed by an order of $10^6$.    
The situation is similar for BP1 and BP3.
}. 
Thus, we will only consider SS-1 and recast the SSDL search by the ATLAS Collaboration~\cite{ATLAS:2018dcj}.

The following selection criteria (\textbf{SR-ee} cuts) are applied in Ref.~\cite{ATLAS:2018dcj}. A pair of same-sign electrons with the transverse momenta $p_T > 30\gev$ and the pseudo-rapidity $|\eta|<2.47$ are selected. All selected events contain at least two jets with $p_T > 100\gev$ and $|\eta|<2.0$. The invariant masses of two electrons $(m_{ee})$ and two jets $(m_{jj})$ satisfy $m_{ee}> 400\gev$ and $m_{jj}> 110\gev$. The scalar sum of $p_T$ of electrons and the two most energetic jets $(H_T)$ is larger than $400\gev$.

In the recast, we simulate the signal process SS-1 and obtain the selection efficiencies passing the \textbf{SR-ee} cuts. Owing to the same selection criteria, the SM backgrounds are the same as those modelled in Ref.~\cite{ATLAS:2018dcj}. Thus we can easily take the number of SM background events from Ref.~\cite{ATLAS:2018dcj} for the integrated luminosity of $36.1\fbi$, which is $11.2$  with the \textbf{SR-ee} cuts being imposed. For a larger integrated luminosity $\mathcal{L}$, the number of SM background events is assumed to scale with the integrated luminosity, that is $11.2\times \mathcal{L}/(36.1\fbi)$.

We use \texttt{MadGraph5\_aMC@NLO}~\cite{Alwall:2014hca} to generate signal events with the PDF set NNPDF3.0NLO~\cite{NNPDF:2014otw}, which are passed to \texttt{Pythia8}~\cite{Sjostrand:2014zea} and \texttt{Delphes3}~\cite{deFavereau:2013fsa} for parton shower and detector simulation. 
The selection efficiencies of signals events in SS-1 passing \textbf{SR-ee} cuts  are 0.30, 0.10, 0.27 and 0.38 for the benchmark points BP1, BP2, BP3 and BP4, respectively~\footnote{They are validated against the efficiencies reported by the ATLAS Collaboration~\cite{ATLAS:2018dcj} in the context of right-handed $W$ boson and heavy neutrino. }. Note that the selection efficiency for BP2 is smaller than those for the other benchmark points, since the energy of electron produced in association with $\Psi$ is limited by the mass difference between $m_S$ and $m_\Psi$.

After imposing \textbf{SR-ee} cuts, a likelihood fit of the $H_T$ distribution was performed in Ref.~\cite{ATLAS:2018dcj} to derive the observed limits, since signals tend to have larger $H_T$ compared to the SM backgrounds. We generate $H_T$ distributions of the signal SS-1 for four benchmark points, which are compared with the SM background $H_T$ distribution in Ref.~\cite{ATLAS:2018dcj}, and find that the cut $H_T > 1.25\tev$ can efficiently separate the signals from the SM backgrounds. The overall selection efficiencies are $0.26$, $0.09$, 0.27 and 0.38 for BP1, BP2, BP3 and BP4, respectively -- which are more-or-less the same as those for $H_T > 400\gev$ -- and only $\sim 1$ background event remains with the integrated luminosity of $36.1\fbi$. While the sensitivity of SSDL search is improved, we will only  impose the \textbf{SR-ee} cuts to avoid an overestimate of the sensitivity.

We evaluate the exclusion limit at 95\% C.L. using the asymptotic
formula~\cite{Cowan:2010js}
\begin{align}
\label{eq:exclusion}
Z_{\text{excl}}\equiv \sqrt{2[s-b\ln{((s+b)/b})]}=1.96\;,
\end{align}
where $s$ and $b$ are the numbers of signal and background events after passing the selection cuts, respectively. 
Besides the expected exclusion limit, we also consider the $5\sigma$ discovery potential for the SSDL search, which is obtained with~\cite{Cowan:2010js}
\begin{align}
\label{eq:discovery}
Z_{\text{disc}}\equiv \sqrt{2[(s+b)\ln{((s+b)/b})-s]}=5\;.
\end{align}
As mentioned earlier, to assess future prospects at the HL-LHC, we assume that the selection cuts and efficiencies remain unchanged.
The sensitivities to the couplings in Eq.~\eqref{eq:interactions} will be discussed in Sec.~\ref{sec:results}.

\subsection{Dijet resonance search}

The mass of the scalar $S$ and its couplings are constrained by the dijet resonance searches at the 13 TeV LHC with the integrated luminosities of $139\fbi$~\cite{ATLAS:2019fgd} and $137\fbi$~\cite{CMS:2019gwf} by the ATLAS and CMS Collaborations, respectively. 
In both searches, the upper limits on product of signal cross section and acceptance, 
the latter of which depends on the transverse momenta and pseudo-rapidities of jets,
for resonances in some benchmark models were obtained. A simple scaling was used in Ref.~\cite{Bordone:2021cca} to reinterpret the results~\cite{ATLAS:2019fgd,CMS:2019gwf} in the benchmark of new gauge boson $W^\prime$ to other mediators. We have verified that the reinterpretation in Ref.~\cite{Bordone:2021cca} works well for  Ref.~\cite{CMS:2019gwf} by the CMS Collaboration. Specifically, we generate the signal $pp\to S^\pm$, and assume that $S^\pm$ only decays into $u\bar{d}$ or $d\bar{u}$, and set $\sigma(pp\to S^\pm)\times A $ to be equal to the limits observed by the CMS Collaboration given in  Ref.~\cite{CMS:2019gwf}, in which the acceptance of $A=0.5$ for a scalar is declared. The upper limit on the coupling of $S^\pm$ to the quarks we obtain agrees well with that in Ref.~\cite{Bordone:2021cca}.
However, the acceptance for a scalar was not reported by the ATLAS Collaboration in Ref.~\cite{ATLAS:2019fgd}. Using the reinterpretation results given in Ref.~\cite{Bordone:2021cca}, we find that the acceptance for the scalar in the ATLAS analysis~\cite{ATLAS:2019fgd}  is about $A=0.60$, $0.89$ and $0.55$ for  $m_S=2\tev$, $3.5\tev$ and $4.5\tev$, respectively.

In Ref.~\cite{ATLAS:2019fgd}, the 95\% C.L. observed upper limits on $\sigma_{{\rm dijet}}\times A$ are obtained after performing the likelihood fit of the dijet invariant mass distribution, where $\sigma_{{\rm dijet}}$ denotes the signal cross section. 
For $m_S=2\tev$, $3.5\tev$ and $4.5\tev$, we obtain that the upper limits on  $\sigma_{{\rm dijet}}$ are $38.3~\text{fb}$, $5.3~\text{fb}$ and $6.3\fb$, respectively.The projection 
of the upper limits are derived using Eq.~\eqref{eq:exclusion} assuming that the numbers of signal and background events both scale with the integrated luminosity. 

The total dijet cross section is given by
\begin{align}
\sigma_{{\rm dijet}} &= \sigma_{S^+}\text{Br}(S^+\to \bar{d}u) + \sigma_{S^-}\text{Br}(S^-\to \bar{u}d)
+ \sigma_{S^0}\text{Br}(S^0\to \bar{d}d)\;
\end{align}
within our signal model, where the contribution of the neutral scalar $S^0$ is also included. Here $\sigma_{S^\pm}$ and $\sigma_{S^0}$ are the production cross sections of $S^\pm$ and $S^0$, respectively, and ``Br'' denotes the decay branching ratio. By requiring the dijet cross section $\sigma_{{\rm dijet}}$ to be smaller than its observed upper limit for a given resonance mass~\cite{ATLAS:2019fgd}, we obtain the exclusion limits of the couplings $y_{qd}$ and $y_{e\Psi}$ at 95\% C.L., which will be presented in Sec.~\ref{sec:results}.

\section{Results and discussion}
\label{sec:results}

In Sec.~\ref{sec:model_DBD} and Sec.~\ref{sec:LHC}, we have investigated the $\onbb$ decay and 
LHC searches in our simplified model, respectively. Our interest is  the regime where both sets of searches provide complementary and/or overlapping sensitivities. Thus, we consider new particle masses that allow for potentially observable effects at the LHC. Specifically,
we assume that the masses $m_S > m_\Psi$ and $m_R = 2\tev$, with benchmark values of $m_\Psi$ and $m_S$  shown in Tab.~\ref{tab:Benchmark}. From the expected sensitivity to $\onbb$-decay rate at future ton-scale experiments, $|y_{e\Psi} y_{qd} \lambda_{ed} \lambda_{u\Psi}|\gtrsim \mathcal{O}(10^{-3})$ is required with a sizable  uncertainty due to unknown values of LECs for vector operators. For illustration, we assume the LECs $g_V^{\pi N}=\tilde g_V^{\pi N} =1$ and $g^{NN}_6=g^{NN}_7=0$, and the couplings $|\lambda_{ed}\lambda_{u\Psi}|=1$. For a larger (smaller) $|\lambda_{ed}\lambda_{u\Psi}|$, the sensitivity of $\onbb$ decay to the couplings $y_{qd}$ and $y_{e\Psi}$ is stronger (weaker),  
while the LHC sensitivities are less affected \footnote{We will not consider the case $|\lambda_{ed,u\Psi}|\ll |y_{e\Psi,qd}|$, which the SSDL and $\onbb$-decay searches are insensitive to.
}. Besides, the sensitivity of $\onbb$ decay becomes stronger (weaker) for  negative (positive) values of $g^{NN}_{6,7}$. 
In order to differentiate the couplings $\lambda_{ed}$ and $\lambda_{u\Psi}$, a more detailed analysis of leptoquark searches in our simplified model is needed. 
As discussed in Sec.~\ref{sec:LHC}, the single production, single resonant production of the leptoquark and the Drell-Yan production are sensitive to the coupling $\lambda_{ed}$.

\begin{figure}[!htb] 
\centering
\includegraphics[width=0.4\linewidth]{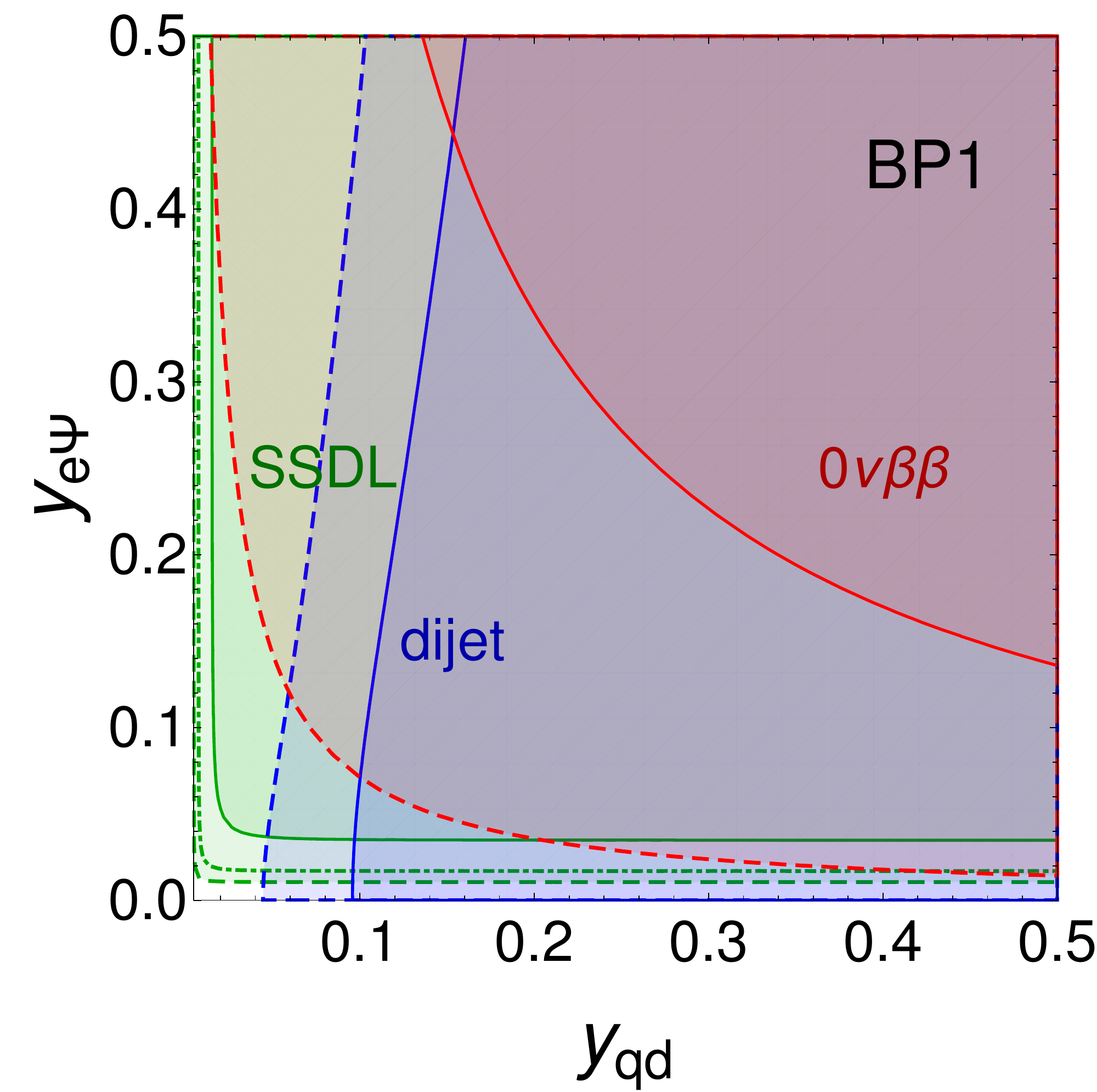}
\includegraphics[width=0.4\linewidth]{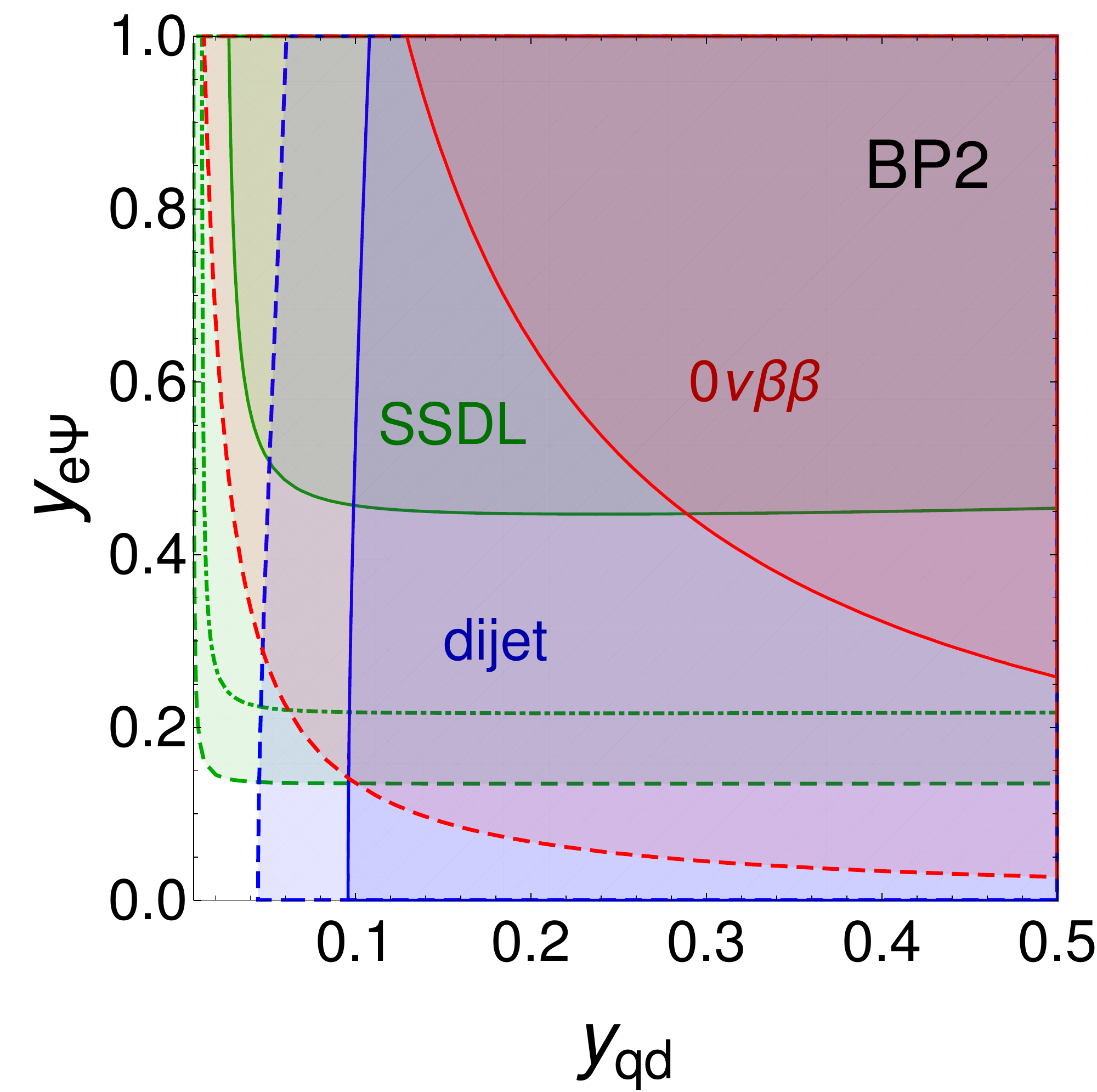}
\caption{The current and projected sensitivities in the plane of $y_{qd}-y_{e\Psi}$
for BP1 and BP2. The red regions are excluded at 90\% C.L. by the KamLAND-Zen (solid curve) and ton-scale (dashed curve) experiments. The blue regions are excluded at 95\% C.L.  by the dijet searches with the integrated luminosities of $139\fbi$ (solid curve) and $3\abi$ (dashed curve). The green regions are excluded at 95\% C.L.  by the SSDL searches  with the integrated luminosities of $36.1\fbi$ (solid curve) and $3\abi$ (dashed curve). The green dot-dashed curves correspond to the $5\sigma$ discovery potential at the HL-LHC. The product of leptoquark couplings $|\lambda_{ed}\lambda_{u\Psi}|=1$ is assumed. $y_{e\Psi}$ and $y_{qd}$ denote the magnitudes of the couplings with the absolute value symbols omitted. 
}
\label{fig:comb-1}
\end{figure}

\begin{figure}[!htb] 
\centering
\includegraphics[width=0.4\linewidth]{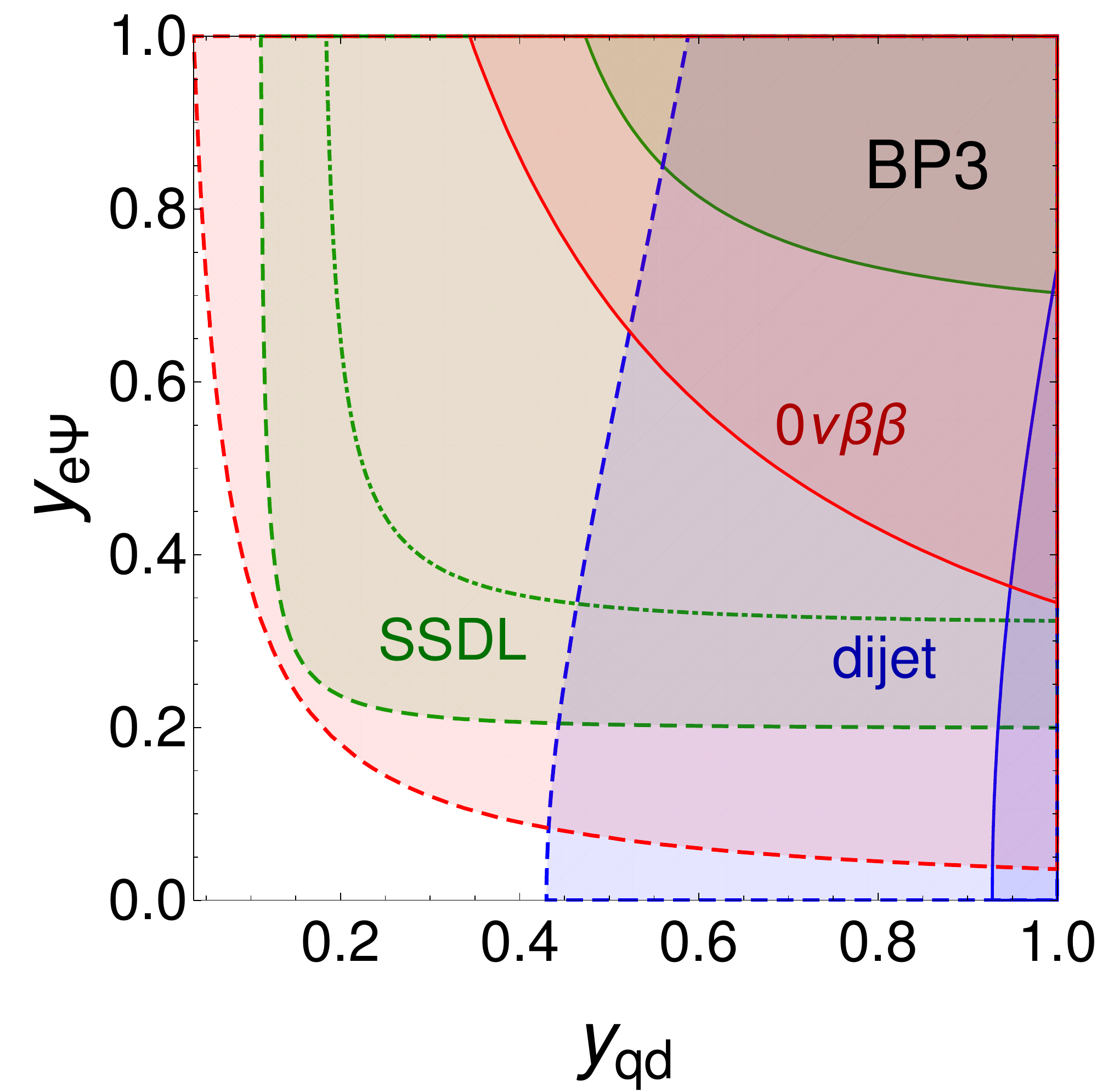}
\includegraphics[width=0.4\linewidth]{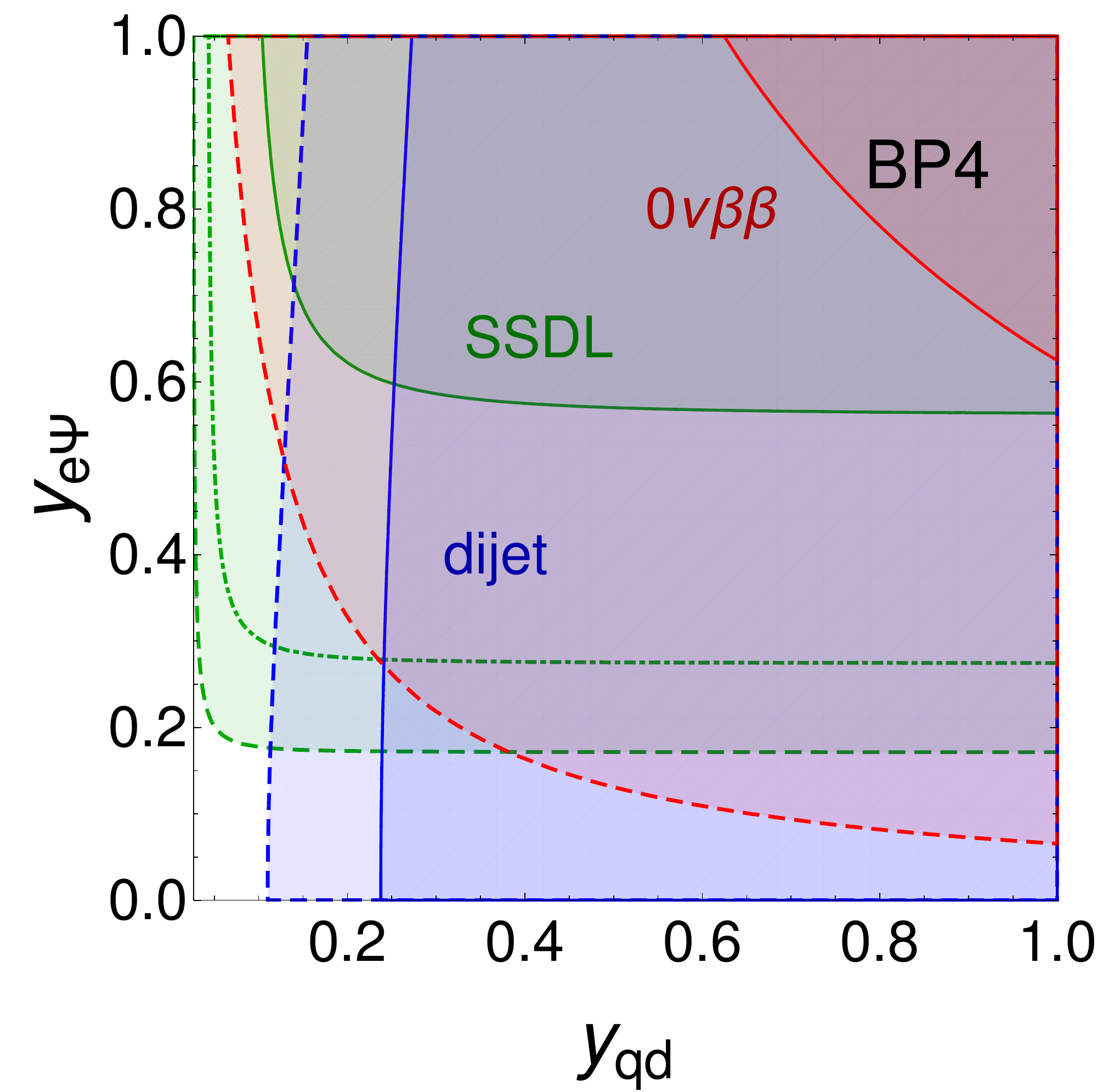}
\caption{Same as Fig.~\ref{fig:comb-1} but for BP3 and BP4.
}
\label{fig:comb-2}
\end{figure}

We show the sensitivities  in the $\onbb$ decay, SSDL search and dijet search in Fig.~\ref{fig:comb-1} and Fig.~\ref{fig:comb-2}. The current and projected exclusion limits are depicted in solid and dashed curves, respectively. 
The red regions are excluded by the $\onbb$-decay experiment KamLAND-Zen and future ton-scale experiments at 90\% C.L. 
The current constraints from the SSDL search (green) and dijet search (blue) are obtained with the integrated luminosities of $36.1\fbi$ and $139\fbi$ at the 13~TeV LHC, respectively, while the future projections for the HL-LHC are made with the integrated luminosity of $3\abi$. 
We also show the $5\sigma$ discovery potential of the SSDL search at the HL-LHC in green dot-dashed curve.

Both the SSDL and dijet searches are sensitive to the scalar mass $m_S$. For relatively small $m_S$, the cross section of $pp\to S^\pm$ is sizable for BP1 and BP2. From the left panel of Fig.~\ref{fig:comb-1}, the accessible region of $y_{qd}$ and $y_{e\Psi}$ (the absolute value symbols are omitted) at future ton-scale $\onbb$-experiments has already been excluded by the existing SSDL and dijet searches for BP1.  In such a scenario, there is no signal expected in $\onbb$-decay experiments barring extremely small $y_{qd}$. 

The $\onbb$-decay and LHC experiments are complementary for the other benchmark points. For BP2, $m_\Psi$ is close to $m_S$, so that the decay branching ratio of  $S^\pm \to e^\pm \Psi$ is largely suppressed and the selection efficiency in the SSDL search is smaller compared to BP1. In the right panel of Fig.~\ref{fig:comb-1}, we can see that the combination of the existing SSDL and dijet searches gives stronger constraints compared to the $\onbb$-decay search at KamLAND-Zen. For $y_{qd}\lesssim 0.1$, which is currently allowed, it is possible to observe a TeV-scale LNV signal in a large portion of the region $ y_{e\Psi}\gtrsim 0.25$ in both future $\onbb$-decay and SSDL searches. If however no signal is observed, most of the region  $y_{e\Psi}\gtrsim 0.15$ can be excluded.

In Fig.~\ref{fig:comb-2}, larger values of $m_S$ are considered. The cross sections of $pp\to S^\pm $ for BP3 and BP4 are smaller than those for BP1 and BP2 by $2-3$ orders of magnitude, as seen in Tab.~\ref{tab:Benchmark}.  For BP3, the sensitivities to the couplings $y_{qd}$ and $y_{e\Psi}$ in KamLAND-Zen and future ton-scale $\onbb$-decay experiments are always better than those of the current and projected SSDL searches, respectively, which is shown in the left panel. In addition, the constraint from the existing dijet search is much weaker than that for BP1 and BP2. For future prospects, the dijet search can probe a large portion of the region $y_{qd}\gtrsim 0.4$. One can observe TeV-scale LNV signal in both future SSDL and $\onbb$-decay searches for $y_{e\Psi} \gtrsim 0.35$ and $y_{qd}\gtrsim 0.15$, which is indicated by the $5\sigma$ contour. 
If signal of LNV is only observed in ton-scale $\onbb$-decay experiments, it implies that either $y_{e\Psi} \lesssim 0.2$ or $y_{qd}\lesssim 0.1$, which could be distinguished using the dijet searches. 

For BP1, BP2 and BP3, we have assumed that the Dirac fermion mass $m_\Psi$ is below the leptoquark mass $m_R$. For BP4, an alternative scenario $m_\Psi > m_R$ is considered, and $m_S$ is slightly larger than $m_\Psi$. In comparison with BP1, the sensitivity of SSDL search is reduced due to the smaller production cross section and decay branching ratio of $S^\pm$, although the decay branching ratio of $\Psi$ increases. Besides, the sensitivity of $\onbb$ decay also degrades, since the decay rate is proportional to $1/(m_S^4 m_R^4 m_\Psi^2)$. In the right panel of Fig.~\ref{fig:comb-2}, we can see that the projected sensitivity of SSDL search at the HL-LHC is better than that of $\onbb$ decay in ton-scale experiments in contrast to BP3, after taking into account the constraint from the dijet search. In analogy with BP2, it is very promising to observe TeV-scale LNV for $y_{e\Psi}\gtrsim 0.3$. We also note that BP2 and BP4 might be  distinguished with differential distributions, such as $p_T$ of electron or jets and $H_T$.

It is worth noting a few caveats about our projections: if the mass of the new particles is sufficiently large, the LHC loses sensitivity and a future $pp$ collider may be needed;
we assumed the same set of selection cuts and efficiencies that occurs, at the time of 
this writing, in 
LHC collaboration analyses of 13 TeV LHC data;
and we did not include any effects due to pile-up or pile-up subtraction. 
Our results  indicate that the HL-LHC has promising prospects for uncovering the chirally-suppressed mechanism with TeV-scale LNV, that is complementary to the on-going $\onbb$ decay experiments. Our results suggest a more detailed study including more realistic assessment of backgrounds and optimizing the reach of the HL-LHC for TeV-scale LNV is warranted.

\section{Conclusion}
\label{sec:conclusion}

In this work, we have studied $\Delta L=2$ LNV interactions, whose contributions to $\onbb$ decay are chirally suppressed in a simplified model.  While the $\onbb$ decay is insensitive to the details of underlying mechanism,
the LHC can provide complementary tests in the same-sign dilepton plus dijet search channel and, 
by using dijet and leptoquark searches, further distinguish different scenarios of TeV-scale LNV in our simplified model. We have calculated the half-life of $\onbb$ decay, and investigated the current and projected sensitivities of the LHC searches.

In three of the benchmark points we examined (BP1, BP2, and BP4), we find that due to the on-shell production of a $2-3.5$~TeV charged scalar $(S^\pm)$ in the $s$-channel at the LHC, the sensitivities of SSDL searches are better than those of $\onbb$-decay searches.
Dijet searches can place severe constraints, and in particular a large portion of parameter space accessible to future SSDL and $\onbb$-decay searches has been excluded by the existing dijet search.
However, for a fourth benchmark point (BP3) in which the scalar particle 
has a mass of 4.5 TeV, we find that $\onbb$-decay searches have a better sensitivity compared to the SSDL search, and the constraint from dijet search is much weaker. In all of the scenarios that we consider, most of 
the region within reach of the $\onbb$-decay search at future ton-scale $\onbb$-decay experiments
can be covered at the $5\sigma$ discovery level in the SSDL search at the HL-LHC.

From the benchmark studies, we could obtain some general results about the future $\onbb$-decay and LHC searches for TeV-scale LNV, which at low energies generates dimension-9 vector operators:
\begin{itemize}
\item If  one observes a LNV signal in the SSDL search but not in $\onbb$ decay, it implies that either the coupling of scalar to quark or lepton is extremely small ({\it e.g.} BP1) or the masses of new Dirac fermion and scalar are close ({\it e.g.} BP2 and BP4). 
\item If one observes a LNV signal in $\onbb$ decay but not in the SSDL search, however, there might exist a heavy scalar with its coupling to quarks or leptons is small ({\it e.g.} BP3). 
\item If LNV signals are observed in both $\onbb$ decay and SSDL search, either the production or decay of the scalar at the LHC is suppressed ({\it e.g.} BP2, BP3 and BP4).
\item  If there is no LNV signal observed, then LNV may have other origins or lepton number may be conserved at the classical (Lagrangian) level.
\end{itemize}
In case of  LNV signal(s) being observed, depending on whether we observe a signal in dijet search, we know that  the coupling of scalar to quarks is large or not.

In this work we have focused on the potential of the LHC to search for the particles and interactions in the simplified model underlie the dimension-9 LNV operators. It would be interesting to extend our analysis of LHC searches. It would also be interesting to revisit the issues explored here in a model that has acceptable levels of flavor-changing neutral current and charged lepton number violating processes.

\begin{acknowledgments}
We would like to thank Daniele S.M. Alves, Yi-Lun Chung, Glen Cowan, Juan Carlos Helo, Xiao-Dong Ma, Emanuele Mereghetti, and Stephane Willocq for very helpful discussions and correspondence. 
The work of MLG is supported by the LDRD program at Los Alamos National Laboratory and by the U.S. Department of Energy, Office of High Energy Physics, under Contract No. DE-AC52-06NA25396. 
The work of SUQ was supported by ANID-PFCHA/DOCTORADO BECAS CHILE/2018-72190146. GL, SUQ, and MJRM were partially funded under the US Department of Energy contract number DE-SC0011095.  GL was also supported in part by the Fundamental Research Funds for the Central Universities, China, Sun Yat-sen University. MJRM was also supported in part under National Science Foundation of China grant number 19Z103010239.
\end{acknowledgments}


\appendix
\allowdisplaybreaks

\section{Hadron-level \tf{$\onbb$}{0vbb}-decay amplitudes: comparing dimension-9 LNV vector and scalar operators}
\label{app:scalar-vs-vector}

Below the GeV scale,
dimension-9 LNV quark-lepton operators  induce local LNV hadron-lepton interactions. 
These can be 
described using chiral pertubration theory, which is an effective theory with an expansion 
parameter $\epsilon_{\chi} \equiv q/\Lambda_\chi$. Here $\Lambda_\chi \simeq 1$ GeV 
and $q$ is a typical momentum-transfer scale of the nuclear $\onbb$ decay $0^+ \rightarrow 0^+$, with $q \simeq m_\pi$.
For dimension-9 LNV operators, their leading interactions with hadrons are through induced 
local $\pi \pi ee$, $\pi N N ee$ and $NNNNee$ operators. 
For the sake of brevity, we will not give the explicit forms of the dimension-9 LNV operators. For further details the reader is referred to Ref.~\cite{Cirigliano:2018yza}.

For all dimension-9 vector operators $O^\mu_{6,7,8,9}$ and $O^{\mu \prime}_{6,7,8,9}$,
no sizable non-derivative $\pi \pi ee$ operators are generated at  LO \cite{Prezeau:2003xn} or through N$^2$LO \cite{Graesser:2016bpz}. 
The LNV operators induced by the vector operators that involve two $\pi$'s are of the form $(\pi \partial^\mu \pi) \overline{e} \gamma^\mu \gamma_5 e^c$, 
and $(\partial^\nu \pi) (\partial_\nu \partial^\mu \pi) \overline{e} \gamma^\mu \gamma_5 e^c$, 
at LO and N$^2$LO respectively.
The point here is that the pion 
equations of motion can be used to 
show that all of these local operators are proportional to the non-derivative operator $\pi \pi ee$ with 
a coefficient proportional to the electron mass $m_e$ and therefore negligible at this level. 

In 
the Weinberg power counting therefore, the leading contributions of vector operators to the amplitude 
arise from $\pi N Nee$ and $NN NNee$ operators which both contribute at $\mathcal{O}(\epsilon^2_\chi)$ to the $nn \rightarrow pp ee$
amplitude.

Na\"ively then, the transition amplitude arising from dimension-9 vector operators is suppressed by $\mathcal{O}(\epsilon^2_\chi)$ compared 
to that of scalar operators that arise from $O_{2,3,4,5}$ and $O^\prime_{2,3}$ which contribute at $\mathcal{O}(\epsilon^0_\chi)$ to the amplitude. \footnote{In this Appendix we do not consider the 
scalar operators $O_1$ and $O^\prime_1$ which give a chirally suppressed contribution to 
$\onbb$ decay.\label{footnote:ignoring-O1-O1prime}}
While we find this expectation to generally be true, it depends on an interplay between
NMEs and the size of unknown LECs that occurs, in the case of scalar operators, 
in the expression for the amplitude. The point of this Appendix is to expand on this statement.

\begin{figure}[!htb] 
\centering
\includegraphics[width=0.35\textwidth]{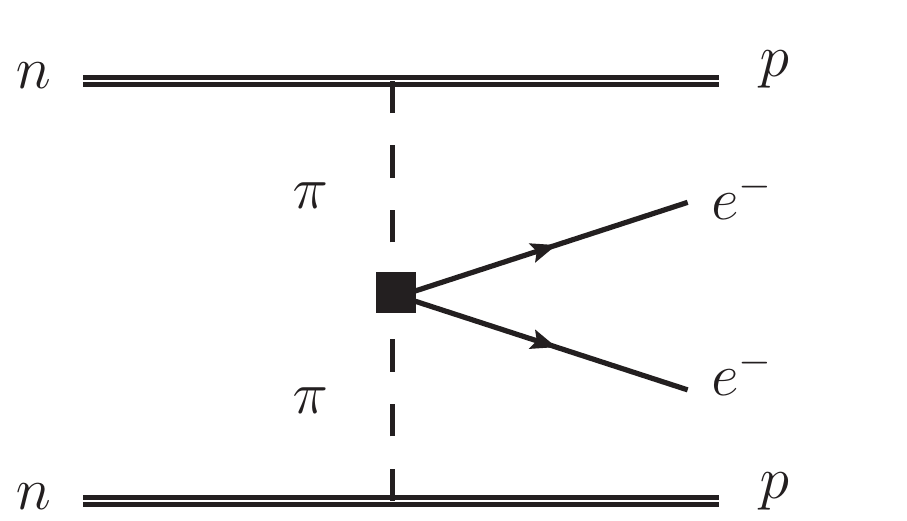}	
\caption{For scalar operators, the amplitude for $nn \rightarrow pp ee$ receives an 
additional contribution from the Feynman diagram shown here, which must be combined 
with those shown in Fig. \ref{fig:feyn_DBD_hadron}.}
\label{fig:feyn_DBD_pipi}
\end{figure}

Going into further detail, the reason dimension-9 scalar 
operators $O_{2,3,4,5}$ and $O^\prime_{2,3}$ contribute at $\mathcal{O}(\epsilon^{0}_\chi)$ 
to the amplitude is because they induce non-derivative $\pi \pi ee$ operators that are unsuppressed by $m_e$ or 
any chiral power. In this case the total amplitude is given by summing the 
Feynman diagrams in Fig. \ref{fig:feyn_DBD_pipi}, together with 
Fig. \ref{fig:feyn_DBD_hadron}. For these scalar operators, in the Weinberg power counting, 
the second diagram in Fig. \ref{fig:feyn_DBD_hadron} is 
na\"ively suppressed to the diagram in Fig. \ref{fig:feyn_DBD_pipi} by $\epsilon^2_{\chi}$. But there is a further 
subtlety: the contribution of the local $\pi \pi ee$ operator to the total amplitude, obtained 
by solving the Lippman-Schwinger equation for the strong $NN$ potential, is actually UV divergent. Requiring that the total amplitude is independent of the regulator necessitates promoting the local $NNNN ee$ operator from N$^2$LO to LO, which would violate Weinberg's power counting. The RG equation relating the LECs of the $\pi \pi ee$ and $NNNNee$ operators suggests 
that the LEC of the latter is actually larger by a factor of $ (4 \pi)^2$ compared to NDA. So that in the chiral 
power counting, the diagram in Fig. \ref{fig:feyn_DBD_pipi} and the second diagram in Fig.~\ref{fig:feyn_DBD_hadron} 
contribute at the same 
order \cite{Cirigliano:2018hja,Cirigliano:2018yza,Cirigliano:2019vdj}. This feature will be important to the discussion that follows.

To see this explicitly, the contributions of dimension-9 scalar operators \footnote{But 
note the previous footnote \ref{footnote:ignoring-O1-O1prime}.}
to the amplitude for nuclear $\onbb$ decay $0^+\to 0^+$ is 
given by \cite{Cirigliano:2018yza}
\begin{align}
\mathcal{A}_{\text{scalar}} = \dfrac{g_A^2 G_F^2 m_e }{\pi R_A} \left[\mathcal{A}_\nu\bar{u}(k_1) P_R  C \bar{u}^T (k_2)  + 
\mathcal{A}_R\bar{u}(k_1) P_L  C \bar{u}^T (k_2)
\right]\;,
\label{eq:amplitude-scalar-operators}
\end{align}
where the reduced amplitudes $\mathcal{A}_\nu$ and $\mathcal{A}_R$ are 
\begin{align}
\mathcal{A}_{\nu} (\mathcal{A}_R)&=& \frac{m^2_N}{m_e v} \left[-\frac{1}{2m^2_N} C^{(9)}_{\pi \pi L(R)} M_{PS,sd} + \frac{m^2_\pi}{2 m^2_N} C^{(9)}
_{\pi N L(R)} M_{P,sd} - \frac{2}{g^2_A} \frac{m^2_\pi}{m^2_N} C^{(9)}_{NNL(R)} M_{F,sd} \right] 
\label{eq:reduced-ampltitude-scalar-op9}
\end{align}
where $C^{(9)}_{\pi \pi L(R)}$, $C^{(9)}_{\pi N L(R)}$, and $C^{(9)}_{NNL(R)}$ are linear in the Wilson coefficients 
$C_i$ of the dimension-9 scalar operators, and in LECs. Their detailed expressions do not matter for this discussion, but can be found in Ref.~\cite{Cirigliano:2018yza}. 
Each of these is multiplied by the NMEs 
\begin{eqnarray} 
M_{PS,sd} &\equiv& \frac{1}{2} M^{AP}_{GT,sd}+M^{PP}_{GT,sd}+\frac{1}{2}M^{AP}_{T,sd}+M^{PP}_{T,sd},  \\
M_{P,sd} &\equiv& M^{AP}_{GT,sd}+M^{AP}_{T,sd}~,
\end{eqnarray}
which are linear combination of short distance 
Gamow-Teller and tensor nuclear matrix elements, and by a short-distance Fermi nuclear matrix element $M_{F,sd}$, respectively. Expressions 
for these matrix elements in terms of ``neutrino potentials'' can also be found in Ref.~\cite{Cirigliano:2018yza}.

Next, first consider the value of $M_{PS,sd}$. 
For $^{136}$Xe, $M_{GT,sd}^{AP}=-2.80$, $M_{GT,sd}^{PP}=1.06$, $M_{T,sd}^{AP}=-0.92$ 
and $M^{PP}_{T,sd}=0.36$, calculated using the quasi-particle random phase approximation (QRPA)~\cite{Hyvarinen:2015bda}. 
While the first three NMEs are $\mathcal{O}(1)$, a partial cancellation occurs in the summation: $M_{PS,sd}=-0.44$. 
A similar pattern of partial cancellation across other isotopes $^{76}$Ge, $^{82}$Se, and $^{130}$Te, whether evaluated by 
QRPA \cite{Hyvarinen:2015bda}, shell models \cite{Menendez:2017fdf}, or interacting boson models (IBM) \cite{Barea:2015kwa,Deppisch:2020ztt} 
is seen to occur: 
\begin{equation}
\begin{tabular}{c|cccc}
$M_{PS,sd}$   & $^{76}$Ge & $^{82}$Se & $^{130}$Te & $^{136}$Xe \\
\hline
\hline
QRPA~\cite{Hyvarinen:2015bda}          & $-0.79$   & $-0.575$  & $-0.78$    & $-0.44$    \\
Shell~\cite{Menendez:2017fdf}         & $-0.315$  & $-0.28$   & $-0.32$    & $-0.25$    \\
IBM ~\cite{Barea:2015kwa}          & $-0.37$   &           &            &      
\end{tabular}
\end{equation}

For vector operators, the NME for the left diagram appearing in Fig. \ref{fig:feyn_DBD_hadron} 
is given by $M_{P,sd}$, 
and is currently 
\begin{equation}
\begin{tabular}{c|cccc}
$M_{P,sd}$   & $^{76}$Ge & $^{82}$Se & $^{130}$Te & $^{136}$Xe \\
\hline
\hline
QRPA~\cite{Hyvarinen:2015bda}          & $-6.2$   & $-4.47$  & $-6.4$    & $-3.7$    \\
Shell~\cite{Menendez:2017fdf}         & $-2.3$  & $-2.1$   & $-2.4$    & $-1.9$    \\
IBM ~\cite{Barea:2015kwa}          & $-2.34$   &           &            &      
\end{tabular}
\end{equation}

Let's now consider each of the three terms in Eq.~\eqref{eq:reduced-ampltitude-scalar-op9} in turn. 
The first one arises from induced $\pi\pi ee$ interactions. Specifically, 
$C^{(9)}_{\pi \pi L(R)}$ depends on the Wilson coefficients $C_i$ and the LECs $g^{\pi\pi}_{i=2,3,4,5}$, 
the latter of which are known from chiral $SU(3)$ \cite{Cirigliano:2017ymo} 
and lattice QCD \cite{Nicholson:2018mwc} to be $\sim $ few $\times$ GeV$^2$, in agreement 
with na\"ive dimensional analysis (NDA). Thus all else being equal, the size of 
the quantity $C^{(9)}_{\pi \pi L(R)}/m^2_N$ is just given by the size of the 
Wilson coefficients $C_i$ appearing in the expressions for $C^{(9)}_{\pi \pi L(R)}$.

For the second term, $C^{(9)}
_{\pi N L(R)}$ depends on $g^{\pi N}_1$ and $g^{\pi \pi}_1$, which at this chiral order only occurs for $O_1$ and $O^\prime_1$.
Since we are not considering $O_1$ and $O^\prime_1$ in this Appendix, we can neglect the second 
term in our comparison of the scalar operators $O_{2,3,4,5}$ and $O^\prime_{2,3}$ to the vector operators. 

The last term in Eq.~\eqref{eq:reduced-ampltitude-scalar-op9} arises from induced 
4-nucleon operators ($NNNNee$), and is from the diagram on the right in Fig. \ref{fig:feyn_DBD_hadron}.  It is 
na\"ively suppressed compared to the first term, depending as it does 
on the explicit factor of $m^2_\pi/m^2_N$. However, the prefactor $C^{(9)}_{NNL(R)}$ depends on the LECs $g^{NN}_i$, and RG evolution suggests 
$g^{NN}_{2,3,4,5} = \mathcal{O}((4 \pi)^2)$ \cite{Cirigliano:2018yza,Cirigliano:2019vdj}. If these LECs are that large then
the last term cannot be neglected. 

But first suppose these LECs are much smaller than  $\mathcal{O}((4 \pi)^2)$. Then the ratio of amplitudes induced 
by 
vector operators to scalar operators is, all else being equal, 
\begin{align}
\frac{\mathcal{A}_{\text{vector}}}{\mathcal{A}_{\text{scalar}}} \simeq \frac{m^2_\pi}{m^2_N}\frac{M_{P,sd}}{M_{PS,sd}}~.
\end{align}
The ratio of NMEs appearing above is currently 
\begin{equation}
\begin{tabular}{c|cccc}
$M_{P,sd}/M_{PS,sd}$   & $^{76}$Ge & $^{82}$Se & $^{130}$Te & $^{136}$Xe \\
\hline
\hline
QRPA          & $7.8$   & $7.8$  & $8.3$    & $8.5$    \\
Shell         & $7.3$  & $7.6$   & $7.6$    & $7.8$    \\
IBM          & $6.3$   &           &            &      
\end{tabular}
\end{equation}
and exhibits remarkable stability across isotopes and methods for estimating the NMEs. In particular, 
this ratio is $\mathcal{O}(6-8)$, and is driven by the ``small'' values of the NME appearing in 
$\mathcal{A}_{\text{scalar}}$, 
rather than a suppression of the NME appearing in $\mathcal{A}_{\text{vector}}$. 

In other words, 
under the assumption 
that the contribution of the right diagram in Fig. \ref{fig:feyn_DBD_hadron} to $\mathcal{A}_{\text{scalar}}$ is 
$\mathcal{O}(m^2_\pi/m_N^2)\simeq 1/60$, the amplitude $\mathcal{A}_{\text{vector}}$ is only suppressed compared to 
$\mathcal{A}_{\text{scalar}}$ by an amount $\simeq (6-8)/60$, rather than $1/60.$

This conclusion is however sensitive to the unknown values of the LECs $g^{NN}_i$ contributing to $\mathcal{A}_{\text{scalar}}$, as we now explain. 
The reason is that in the expression for $\mathcal{A}_{\text{scalar}}$, the contribution of the 4-nucleon 
operator is weighted by the NME $M_{F,sd}$, which are not small: 
\begin{equation}
\begin{tabular}{c|cccc}
$M_{F,sd}$   & $^{76}$Ge & $^{82}$Se & $^{130}$Te & $^{136}$Xe \\
\hline
\hline
QRPA~\cite{Hyvarinen:2015bda}          & $-3.46$   & $-2.53$  & $-2.97$    & $-1.53$    \\
Shell~\cite{Menendez:2017fdf}         & $-1.46$  & $-1.37$   & $-1.61$    & $-1.28$    \\
IBM ~\cite{Barea:2015kwa}          & $-1.1$   &           &            &      
\end{tabular}
\end{equation}

In the extreme situation that the LECs $g^{NN}_{2,3,4,5} \simeq \mathcal{O}((4 \pi)^2)$, the relevant comparison of NME's is instead
to $M_{P,sd}/M_{F,sd}$ 
\begin{equation}
\begin{tabular}{c|cccc}
$M_{P,sd}/M_{F,sd}$   & $^{76}$Ge & $^{82}$Se & $^{130}$Te & $^{136}$Xe \\
\hline
\hline
QRPA~\cite{Hyvarinen:2015bda}          & $1.8$   & $1.8$  & $2.2$    & $2.4$    \\
Shell~\cite{Menendez:2017fdf}         & $1.6$  & $1.5$   & $1.5$    & $1.5$    \\
IBM ~\cite{Barea:2015kwa}          & $2.1$   &           &            &      
\end{tabular}
\end{equation}
which varies on the $\mathcal{O}(1.5-2.5)$.

Of course, 
the enhancement of the LECs appearing in the $\onbb$-decay amplitude induced by the 
scalar operators $O_{2,3,4,5}$ and $O^\prime_{2,3}$ may not be as large as $(4 \pi)^2$. 
In that case the comparison between the $\onbb$-decay 
amplitude induced by vector and scalar operators lies somewhere between the two extremes 
described here. Reducing the uncertainty on the sizes of the LECs is greatly needed.

\section{Neutrino mass}
\label{app:neutrino_mass}

\begin{figure}[!htb] 
\centering
\includegraphics[width=0.65\textwidth]{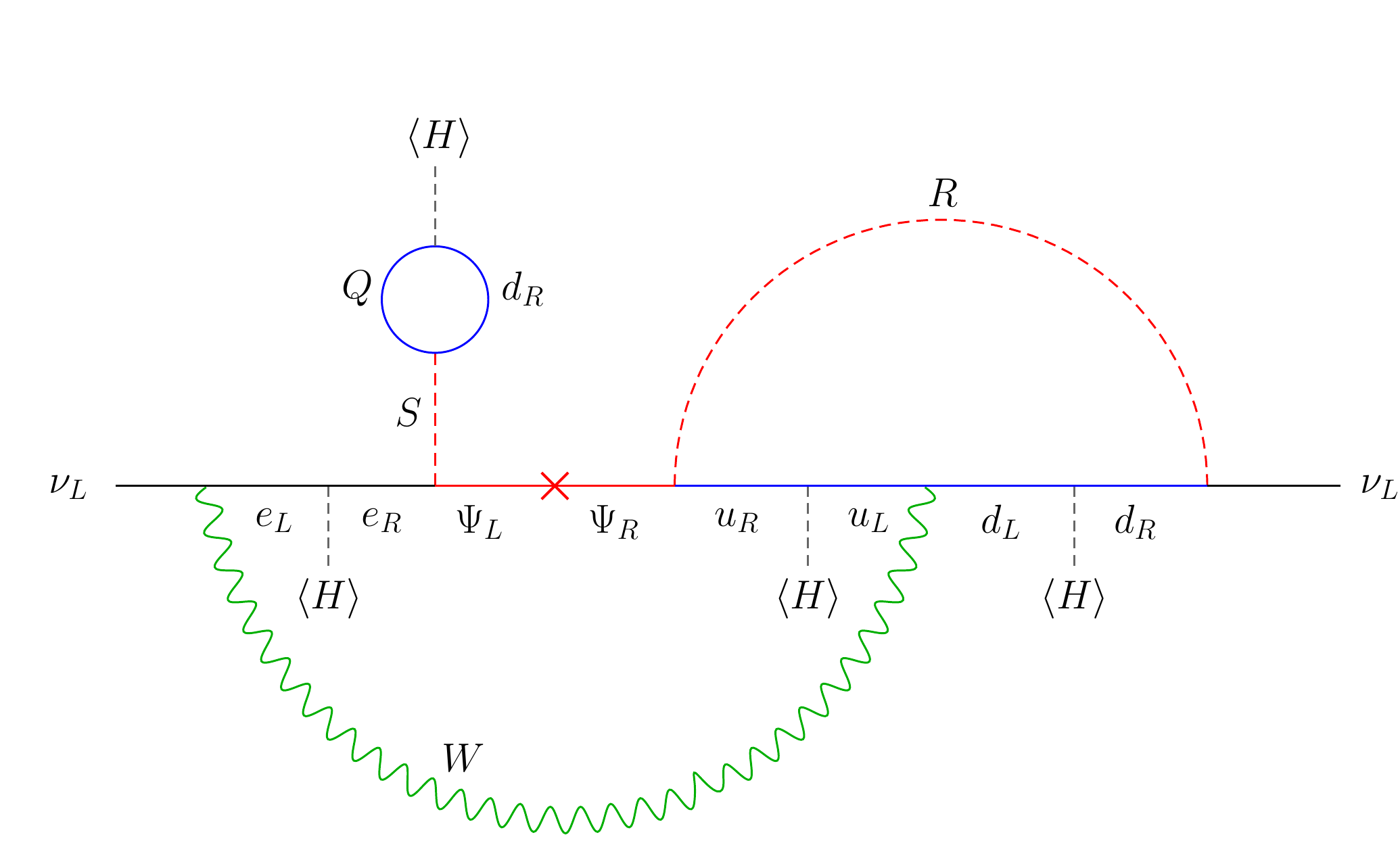}	
\caption{An illustrative LNV Feynman diagram of the UV theory that contributes to the neutrino mass. The heavy degrees of freedom are highlighted in red.}
\label{fig:neutrino_mass_full}
\end{figure}

In the full theory where all the fields are not integrated out, a neutrino mass is first generated at three-loop order, 
as shown, for example, by the Feynman diagram given in Fig.~\ref{fig:neutrino_mass_full}.
If $S$ acquires a vev $\langle S \rangle$, then a neutrino mass is generated at two-loop order, as 
can be seen from a straightforward inspection of Fig.~\ref{fig:neutrino_mass_full}.
In either case, the estimate size of the neutrino mass from these effects is far below the actual value of 
neutrino masses, 
as discussed in Sec.~\ref{sec:model_DBD}.

\section{Partial decay widths}
\label{app:prod_decay}

In this appendix, we provide the decay widths of the new particles in our simplified model. For brevity, all of the couplings are assumed to be real and the step functions $\theta(m_X-m_Y)$ for the decay processes $X\to Y+\dots$  are omitted.

The two-body decay widths of $\Psi^0$ and $S^+$ are 
\begin{align}
\label{eq:S_two-body}
\Gamma(S^+ \to \Psi^0 e^+) &= \dfrac{y_{e\Psi}^2 (m_S^2-m_\Psi^2)^2 }{16\pi m_S^3}\;,
\end{align}
\begin{align}
\Gamma(\Psi^0 \to S^+ e^-) &= \dfrac{y_{e\Psi}^2 (m_\Psi^2-m_S^2)^2 }{32\pi m_\Psi^3}\;,  
\end{align}
\begin{align}
\Gamma(S^+\to \bar{d}u) &= \dfrac{3}{16\pi} m_S (y_{qu}^2+y_{qd}^2)\;,
\end{align}
\begin{align}
\Gamma(\Psi^0 \to {R}^{2/3}  \bar{u})& = \Gamma( \Psi^{-} \to {R}^{-1/3} \bar{u})\nn\\
&= \dfrac{\lambda_{u\Psi}^2 (m_\Psi^2-m_R^2)^2}{16\pi m_R^3}\;.
\end{align}
The two-body decay widths of $\Psi^-$ and $S^0$ are
\begin{align}
\Gamma(S^0\to \bar{d}d) &= \dfrac{3}{16\pi} m_S y_{qd}^2\;,
\end{align}
\begin{align}
\Gamma(S^0\to \bar{u}u) &= \dfrac{3}{16\pi} m_S y_{qu}^2\;,
\end{align}
\begin{align}
\Gamma(\Psi^- \to S^0 e^-) &= \dfrac{y_{e\Psi}^2 (m_\Psi^2-m_S^2)^2 }{64\pi m_\Psi^3}\;,
\end{align}
\begin{align}
\Gamma(S^0 \to \Psi^+ e^-) &=  \dfrac{y_{e\Psi}^2 (m_S^2-m_\Psi^2)^2 }{32\pi m_S^3}\;,
\end{align}
\begin{align}
\Gamma(S^0 \to \Psi^- e^+) &=  \dfrac{y_{e\Psi}^2 (m_S^2-m_\Psi^2)^2 }{32\pi m_S^3}\;.
\end{align}

The two-body decay widths of ${R}$ are
\begin{align}
\Gamma({R}^{2/3} \to e^+ d) &= \Gamma({R}^{-1/3} \to \bar{\nu} d)\nn\\
& = \dfrac{\lambda_{ed}^2 m_R}{16\pi}\;,
\end{align}
\begin{align}
\Gamma({R}^{2/3} \to \Psi^0 u)& = \Gamma({R}^{-1/3} \to \Psi^{-} u)\nn\\ 
&= \dfrac{\lambda_{u\Psi}^2 (m_R^2-m_\Psi^2)^2}{16\pi m_R^3}\;.
\end{align}

For the three-body decay widths of $\Psi^0$ are computed numerically using \texttt{MadGraph5\_aMC@NLO}~\cite{Alwall:2014hca} and checked with \texttt{CalcHEP}~\cite{Belyaev:2012qa}. For the sake of convenience, we provide the decay widths in the limit of $m_\Psi\ll m_S$, $m_R$
\begin{align}
\Gamma(\Psi^0 \to e^- u \bar{d}) 
& = \dfrac{1}{2048\pi^3} \dfrac{m_\Psi^5}{m_S^4} y_{e\Psi}^2 (y_{qu}^2+y_{qd}^2)\;,
\end{align}
\begin{align}
\Gamma(\Psi^0 \to e^+ \bar{u} d) 
& = \dfrac{1}{2048\pi^3} \dfrac{m_\Psi^5}{m_R^4} \lambda_{ed}^2\lambda_{u\Psi}^2\;.
\end{align}
Because the lepton number of $\Psi^0$ is $+1$, the decay $\Psi^0 \to e^- u \bar{d}$ conserves lepton number 
whereas $\Psi^0 \to e^+ \bar{u} d$ violates it.

\newpage

\bibliographystyle{JHEP}
\bibliography{reference}

\end{document}